\documentclass[aps,pre,twocolumn,groupedaddress,showpacs,preprintnumbers,amsmath,amssymb]{revtex4}
\usepackage{graphics,graphicx} 

\newcommand{\Tr}{\operatorname{Tr}}

\newcommand{\meanval}[1]{\left\langle #1 \right\rangle} 
\newcommand{\InsertFig}[1]{\includegraphics{#1.eps}} 

\begin{document}

\title{Reexamination of the long-range Potts model: A multicanonical approach}
\author{S. Reynal}
\email{reynal@ensea.fr}
\homepage{http://www.ensea.fr/staff/reynal}
\altaffiliation{Permanent address: ENSEA, 6 Av. du Ponceau, 95014 Cergy Cedex, France.}
\author{H.T.Diep}
\affiliation{Laboratoire de Physique Th\'eorique et Mod\'elisation, CNRS-Universit\'e de Cergy-Pontoise, 5 mail Gay-Lussac, Neuville sur Oise, 95031 Cergy-Pontoise Cedex, France}

\begin{abstract}
We investigate the critical behavior of the one-dimensional $q$-state Potts model with long-range (LR) interaction
$1/r^{d+\sigma}$, using a multicanonical algorithm.  
The recursion scheme initially proposed by Berg is improved so as to make
it suitable for a large class of LR models with unequally spaced energy levels. 
The choice of an efficient predictor and a reliable convergence criterion is discussed.
We obtain transition temperatures in the first-order regime which are in far better agreement 
with mean-field predictions than in previous Monte Carlo studies. 
By relying on the location of spinodal points and resorting to scaling arguments, we determine
the threshold value $\sigma_c(q)$ separating the first- and second-order regimes to two-digit precision
within the range $3 \leq q \leq 9$.
We offer convincing numerical evidence supporting $\sigma_c(q)<1.0$ for all $q$, by virtue of
an unusual finite-size effect, namely, 
 finite-size scaling predicts a continuous transition in the thermodynamic limit, 
 despite the first-order nature of the transition at finite size. 
 A qualitative account in terms of correlation lengths is provided.
Finally, we find the crossover between the LR and short-range regimes to occur inside a narrow window
$1.0 < \sigma < 1.2$, thus lending strong support to Sak's scenario.
\end{abstract}

\pacs{05.10.Ln, 64.60.Cn, 75.10.Hk}

\maketitle

\section{Introduction}

Microscopic models with long-range (LR) interactions decaying as a power law, i.e., as $1/r^{d+\sigma}$, have aroused
renewed interest during the last decade. 
Beyond their fundamental relevance to the understanding of critical phenomena, they have started playing a seminal role
in the modeling of neural networks \cite{Amit1989} and spin glasses with Ruderman-Kittel-Kasuya-Yosida (RKKY) interactions \cite{Ford1982},
systems undergoing phase separation, e.g., 
highly ionic systems \cite{PitzerLimaSchreiber1985} and
model alloys \cite{GiacominLebowitz1996},
and more widely in a large class of chemical or biological models where electrostatic interactions, 
polarization, or van der Waals forces play a central role.
They have also attracted much attention in the framework of nonextensive thermodynamics,
where a possible equivalence with short-ranged (SR) models is under consideration \cite{Tsallis1995}. 
 
Since the very early work of Ruelle \cite{Ruelle1968}, LR spin models in particular have been extensively studied.
In one-dimensional models, it has been widely shown that long-range
order occurs at finite temperature if and only if $\sigma\leq 1$ 
\cite{Ruelle1968,Thouless1969,Dyson1969,SimonSokal1981,FrohlichSpencer1982}, and this is in strong contrast to
 the SR case where no phase transition
 exists at finite temperature. 
Fisher and co-workers \cite{FisherMaNickel1972} have shown that the upper critical dimension is reduced to $d^*=2 \sigma$, 
whereby one-dimensional LR models exhibit mean-field-like behavior for $\sigma<0.5$, with
the critical exponents taking on their classical values $\nu=1/\sigma$ and $\gamma=1$ provided the phase transition is continuous. 
Conversely, the critical behavior for $\sigma\geq 0.5$ yields nontrivial exponents, and
 LR models in effect go through a variety of universality classes as $\sigma$ is varied within this range, 
thus exhibiting rich critical behavior.
 Due to the ability to continuously vary the range of interaction, 
 which in effect alters the effective dimension
 of the model, one-dimensional LR models are therefore 
 a powerful paradigm for studying the dependence of critical properties on dimensionality, 
 e.g., in systems above their upper critical dimension \cite{LuijtenBlote1996}.

While significant emphasis has been placed on the Ising chain (see, e.g., \cite{BinderLuijten2001} for a review),
specific studies of the LR $q$-state Potts model are less numerous and rather recent.
These include a transfer matrix study combined with finite-range scaling (FRS) \cite{GlumacUzelac1993},
renormalization group (RG) analyses based on Wilson's momentum-shell method \cite{PriestLubensky1976,TheumannGusmao1985} 
or a real-space procedure \cite{Cardy1981,CannasMagalhaes1997},
a cluster mean-field approach \cite{Monroe1999}, 
and Monte Carlo (MC) simulations \cite{GlumacUzelac1997_1998,GlumacUzelac1999,BayongDiepDotsenko1999,GlumacUzelac2000,KrechLuijten2000,LuijtenMessingfeld2001}.
The last, however, mostly focused on the case $q=3$, 
and led to numerical estimates of critical exponents and temperatures showing some discrepancies. 
Due to the higher ground state degeneracy, this model reveals a phase diagram markedly richer than that of the Ising chain. 
It has been shown in the SR case that the transition turns from a continuous to a first-order
one as the number of states $q$ is increased beyond a threshold value $q_c(d)$ depending on the dimensionality of the model.
For instance, $q_c(2)=4$ and $q_c(4)=2$ \cite{AharonyPytte1981,Baxter1973} (see also Ref.~\cite{Wu1982} for a complete review).
As for the LR case in $d=1$, Glumac and Uzelac have shown from MC studies of the three- and five-state Potts model \cite{GlumacUzelac1997_1998} 
that the same sort of behavior occurs, i.e., there is a so-called tricritical point at some value $\sigma_c(q)$ depending on $q$, 
and the transition is continuous for $\sigma > \sigma_c$. 
This qualitative picture was later reinforced in \cite{BayongDiepDotsenko1999} for $q=3,5,7,9$ 
and in \cite{KrechLuijten2000} for $q=3$, both relying on MC studies, and in \cite{GlumacUzelac1999} using
a graph-weights approach. 
On the other hand, it is noteworthy that RG analyses dedicated to LR models
have remained thus far rather inconclusive, where distinguishing
between first- and second-order transitions is concerned \cite{PriestLubensky1976,TheumannGusmao1985,Cardy1981,CannasMagalhaes1997}.

Although it is now believed that $q_c$ depends continuously on the range of interaction for this class of models, 
the exact location of the tricritical line separating both regions is still fairly controversial. 
The biggest hurdle for a precise and reliable determination of this borderline actually stems from
 the weakening of the discontinuous transition as $\sigma_c$ is approached, even if exceptionally large lattice sizes 
 are simulated, e.g., using the efficient Luijten-Bl\"ote cluster algorithm \cite{LuijtenBlote1995}.
While for $q=3$ $\sigma_c$ was claimed to lie between $0.6$ and $0.7$ in \cite{GlumacUzelac1997_1998}, Krech and Luijten
pointed out that $\sigma=0.7$  still belongs to the first-order regime, and that the second-order regime may set in for 
$\sigma=0.75$ only \cite{KrechLuijten2000}. The situation with $q=5$ turns out to be even worse, with numerical estimates available only
within fairly large ranges: a lower boundary value of $0.8$ was reported in \cite{GlumacUzelac1997_1998},
 whereas $0.7 < \sigma_c(5) < 1.0$ according to \cite{BayongDiepDotsenko1999}. These results have not yielded a
very precise phase diagram as yet, with the only reliable assertion being that $\sigma_c(q)$ increases with $q$.

The marginal case $\sigma=1$ raises another set of thorny questions:  
in \cite{BayongDiepDotsenko1999} it was reported that the
phase transition changes from a second-order to a first-order one for $q\geq 9$, 
while it has been shown by Kosterlitz, using a model with a continuum of states \cite{Kosterlitz1976}, 
and later on by Cardy, using a discrete model \cite{Cardy1981}, 
that inverse square interactions give rise to a Kosterlitz-Thouless (KT) transition,
i.e., one governed by topological defects \cite{Kosterlitz1974}.
It is worth mentioning that both hypotheses may be reconciled, at least partially, 
 by following a scenario similar to the one devised in 
 \cite{DomanyMukamelSchwimmer1980,DomanySchickSwendsen1984,VanEnterShlosman2002}, 
whereby for $XY$-like models with nonlinear nearest-neighbor interactions,
 the KT-like transition is preempted by a first-order transition whenever the nonlinearity becomes 
 strong enough.
While the recent work of Luijten and Messingfeld on the three-state Potts model \cite{LuijtenMessingfeld2001}  
lends further support to Cardy's assertion, 
the controversy still appears unsettled, however, and in this view 
a determination of the asymptotic behavior of $\sigma_c(q)$ as $q \rightarrow \infty$ seems of major interest indeed.

We wish to shed light on some of these contradictory results using MC simulations in
generalized ensembles, with particular emphasis put on the first-order regime.
The aim of this work is thus twofold.
First, we propose an implementation of the multicanonical algorithm dedicated to the numerical study of LR models.
This algorithm, devised by Berg and Neuhaus a decade ago \cite{BergNeuhaus1991,Berg1992}, 
has been successfully applied in the past to SR models undergoing first-order transitions.
As numerical studies of models exhibiting first-order transitions are dramatically hampered 
by huge tunneling times when using standard Metropolis update mechanisms \cite{BergNeuhaus1992,JankeKappler1995}, a multicanonical approach is
indeed an appropriate choice for both the determination of the location of the tricritical line and
 the estimation of critical couplings in the first-order region $\sigma < \sigma_c$. 
 Our purpose is therefore to adapt the scheme initially proposed for SR models so as to make it suitable for a large class of LR models. 
Second, by relying on an extensive study for $3\leq q \leq 9$ and a wide range of $\sigma$ values, 
we arrive at convincing conclusions regarding the location of the tricritical line, the range of validity
of the mean-field-like behavior, which we find much larger than in previous studies, 
and the crossover from the LR to the SR regime, although the last was investigated
for the three-state model only. We show that our multicanonical implementation yields numerical estimates 
 which are in agreement with and often better than those found in previous studies, although our simulations were
performed by relying on medium lattice sizes, i.e., $L \leq 400$ spins.  
In particular, we
obtain the following estimates for $\sigma_c(q)$: $\sigma(3)=0.72(1)$,$\sigma(5)=0.88(2)$,
$\sigma(7)=0.94(2)$ and $\sigma(9)=0.965(20)$, and these results are highly precise.
We also offer convincing evidence that the phase transition in the limiting case $\sigma=1.0$ is not of the first order
for all values of $q$, by virtue of an unusual finite-size effect.
A detailed finite-size scaling (FSS) analysis conducted for $q=9$ shows
that, while the transition belongs to the first-order regime at finite lattice size,
 its first-order nature wanes quickly enough as size is increased so that the transition
tends to a continuous one in the thermodynamic limit.
We give a qualitative account of this behavior in terms of correlation lengths,
and by raising some open questions regarding the dynamics of first-order
transitions in the LR case, we try to challenge the usual picture inherited from SR models. 
Finally, by relying on the shape of the specific heat and computing several moments of the magnetization, 
we conclude that a crossover between LR and SR regimes occurs inside a narrow window $1.0 < \sigma < 1.2$ 
  
The layout of this article is as follows.
In Sec.~\ref{sec:model_theory}, we first review some prominent features 
of the LR Potts model through a mean-field (MF) analysis. Special emphasis is given to the calculation of the location of spinodal points,
 a feature we will use in Sec.~\ref{sec:results} for estimating $\sigma_c(q)$.
Section~\ref{sec:muca_algo} is devoted to implementation details of the multicanonical algorithm specific to LR models. 
We discuss the iteration procedure used to obtain the best estimate for the density of states, 
 the choice of an efficient predictor, and a reliable convergence criterion. 
 Improvements over the original algorithm
are made in order to work out the algorithm instability due to low energy levels being unequally spaced.
Numerical results regarding both first- and second-order regimes are then presented in Sec.~\ref{sec:results}.
Since we do not know of any previous implementation of a generalized ensemble algorithm in the
case of LR spin models, we pay particular attention to comparison with other standard MC algorithms,
i.e., in terms of dynamical exponents, tunneling times, and accuracy of numerical estimates of critical couplings. 
 
\section{Model and mean-field theory} \label{sec:model_theory}

Throughout this work we consider  a ferromagnetic Potts model incorporating LR interactions in $d=1$.
This model is derived from a generalized $q$-state Potts Hamiltonian, i.e.,
$$
 H = -\frac{1}{2} \sum_{i\neq j} J_{ij} \delta_{\sigma_i,\sigma_j}
     - \sum_i h_i \delta_{\sigma_i,\sigma_0},
$$
where the Potts spin $\sigma_i$ at site $i$ can take on the values 
$1,\ldots,q$, the first sum runs over all pairs of sites, and $h_i$ is an
external aligning field favoring condensation in state $\sigma_0$. 
Incorporation of LR interactions is carried out by setting 
$$
	J_{ij} = J(|i-j|)=\frac{1}{|i-j|^{d+\sigma}},
$$
where $d=1$ throughout this study, and $\sigma$ is an adjustable parameter which
can be related to the effective dimension of the model.
As $\sigma$ falls off to $-1$, this model tends to the mean-field case where all 
interactions have equal strength, whereas the limiting case $\sigma \rightarrow \infty$
corresponds to a pure SR model. Crossover from LR to SR behavior should
actually take place at $\sigma=1.0$ \cite{Sak1973,WraggGehring1990}, yet no numerical evidence has been given so far 
for this model which would reinforce this assertion. 
The thermodynamics of the model is studied numerically by way of the following order
parameter:
$$
	M = \frac{q \max_n \rho_n -1}{q-1},
$$
where $n=1,\ldots,q$, and $\rho_n$ is the density of Potts spins in state $n$,
which varies between $1/q$ at infinite temperature and $1$ in the ground state.
On a lattice of size $L$, numerical implementation is carried out by using periodic boundary conditions,
i.e., one adds up interactions between all the spins of the original lattice only, and replaces 
 the bare coupling constant $J(r)$ 
 by $\tilde{J}(r)=\sum_{n=-\infty}^{+\infty} J(r+nL)$.
Retaining only interactions such that $|i-j|<L/2$ 
leads indeed to strong shifts in energy and critical couplings for low $\sigma$ values, especially when
finite-size scaling is to be used with medium lattice sizes. 
For the purpose of numerical evaluation, this sum may be reexpressed as
\begin{align*}
\tilde{J}(r) = \frac{1}{r^{1 + \sigma }} 
		&+ \frac{1}{L^{1 + \sigma }} \, \left[ \zeta\left(1 + \sigma ,1+\frac{r}{L}\right) \right.\\ 
	&+ \left.\zeta\left(1 + \sigma ,1-\frac{r}{L}\right) \right],
\end{align*}
where $\zeta(s,\alpha)$ denotes the Hurwitz zeta function. 
The self-energy will be omitted since it is just an additive constant to the total energy.

Mean-field behavior can be readily obtained by using a variational MF 
approach (see, for instance, \cite{ChaikinLubenskyBook1995}), which relies on the minimization of the following
 functional 
$$
	F[\rho] = \Tr \rho H + kT \Tr\rho\ln\rho
$$
with respect to a trial density matrix $\rho$. Here the trace operation means a sum
over all spin configurations, and the dependence of $H$ and $\rho$ on the spin
configuration is implied. $F[\rho]$ reaches a minimum 
whenever $\rho = e^{-H/kT}/Z$, i.e., in the case of a canonical Gibbs distribution,
and this minimum yields the free energy of the system.
The mean-field approximation allows
us to express the density matrix $\rho$ of the whole system 
as a product of one-site density
matrices $\rho_i$ which depend solely on the spin variable at site $i$. 
We may thus rewrite the trace operation as a sum involving
traces on single spin variables, namely,
\begin{align*}
	F[\rho] &= kT \sum_i \Tr_i \rho_i \ln\rho_i 
			 -\sum_i h_i \Tr_i \rho_i \delta_{\sigma_i,\sigma_0}	\\
	& -\frac{1}{2} \sum_{i\neq j} \Tr_i \rho_i \Tr_j  \rho_j J_{ij} \delta_{\sigma_i,\sigma_j}.
\end{align*}
For further comparison with numerical results, we are mainly interested in expressing
 the free energy as a function of an order parameter which is as similar as possible to the one defined above.
 This is carried out by parametrizing the trial density matrix $\rho_i$ in terms of the following order parameter field:
$$
	m_i = \meanval{\frac{q\delta_{\sigma_i,\sigma_0}-1}{q-1}}_{\rho_i},
$$
where the average is weighted by the trial density matrix $\rho_i$.
Seeing that all states but state $\sigma_0$ are equivalent, 
the constraint $\Tr \rho_i=1$ thus yields 
$$
	\rho_i(m_i,\sigma_i) = \frac{1-m_i}{q}+m_i \delta_{\sigma_i,\sigma_0}.
$$
Considering a uniform external field $h_i=h$, we have $m_i=m$ for all sites;
hence the free energy per spin $f(m)$ reduces to
\begin{align}\label{equ:free_energy_MF}
	\frac{q f(m)}{q-1}  =  	&-  h m - \zeta(1+\sigma) m^2 + kT \{(1-m)\ln(1-m) \nonumber \\ 
				&+ \frac{1+m(q-1)}{q-1}\ln[1+m(q-1)] \}
\end{align}
where we dropped terms which are constant in $m$ so that $f(0)=0$, and
 $\zeta(1+\sigma)$ is the Riemann zeta function. 
This function expands as $1/\sigma$ around $\sigma=0$; hence transition temperatures
are expected to vary as $1/\sigma$ in the vicinity of the MF regime.
Equilibrium values of the order parameter are located at minima of the free energy, 
and it can be seen that $m=0$ is a stable minimum for $kT \geq 2 \zeta(1+\sigma)/q$. 
For $q=2$, there is no third-order term in the series expansion of $f(m)$; hence a second-order transition occurs
at $kT_c = \zeta(1+\sigma)$. 
For $q\geq 3$, the negative coefficient in the third-order term of the series expansion creates a
second minimum, which physically corresponds to a first-order transition. At the transition
temperature, the free energy has the same value at both minima.  
Following \cite{Wu1982}, the exact transition temperature $kT_c$ may be computed by simultaneously solving
 $f(m)=f'(m)=0$ and yields 
$$\frac{kT_c}{\zeta(1+\sigma)} = \frac{q-2}{(q-1)\,\ln(q-1)}.$$ 

\begin{figure}[bt]
	\centering
	\InsertFig{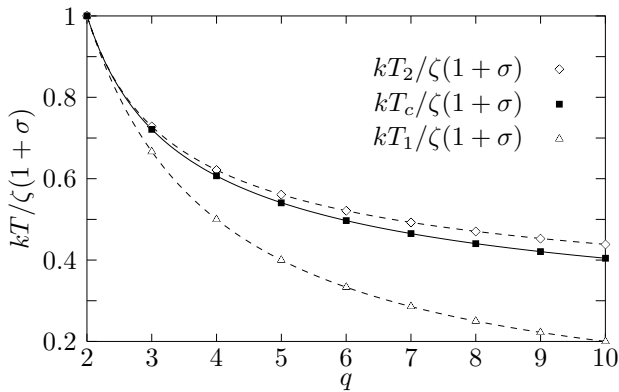}
	\caption{Reduced temperatures of spinodal points $kT_1/\zeta(1+\sigma)$ and $kT_2/\zeta(1+\sigma)$ 
	together with the reduced transition temperature $kT_c/\zeta(1+\sigma)$,
	as a function of $q$ in the MF approximation.}
	\label{fig:theory_potts_model_metastable_states}
\end{figure}
Similarly, spinodal points are computed by jointly solving $f'(m)=f''(m)=0$, giving temperature points at
which either one of the two minima vanishes. These equations possess one trivial solution, namely, 
$kT_1=2\zeta(1+\sigma)/q$ corresponding to the extrema at $m=0$ becoming unstable, and a nontrivial
solution $kT_2$ which may be obtained numerically by solving the following equation, 
$$
\frac{K}{2} \frac{qS-2}{q-1} = \ln \left( S \sqrt{\frac{Kq}{2}} \right),
$$
where $S=1+\sqrt{1+2(1-q)/(Kq)}$
and we have set $K=\zeta(1+\sigma)/kT_2$. Alternatively, one may also express $f$ as a function of 
the MF energy $E=-\zeta(1+\sigma) m^2$ and impose $f'(E)=f''(E)=0$. While these equations yield the
same $kT_1$ and $kT_2$ as above, the two expressions of $f$ obviously do not have the same shape. 
Spinodal points are sketched in Fig.~\ref{fig:theory_potts_model_metastable_states}
for $q$ between $2$ and $10$. These correspond to the limit of metastability for
each subphase, respectively. For temperature points lying inside this temperature range, there exist
two values of the order parameter corresponding to a null curvature of the
free energy, a feature which is known to induce a long-ranged (i.e., low wave number) instability. This in turn triggers
a phase transition through the so-called spinodal decomposition \cite{DombLebowitz1989}.
As expected, the width of the spinodal curve $T_2-T_1$ shrinks to zero as $q\rightarrow 2$, and 
 accounts for the second-order nature of the transition at $q=2$, since
in this limit the two minima merge into a single large minimum responsible for the 
well-known divergence of fluctuations at a continuous transition.
 
\section{The multicanonical algorithm} \label{sec:muca_algo}

The Metropolis algorithm (hereafter denoted as belonging to the class of \textit{canonical} algorithms, i.e., relying on a Boltzmann weighting) 
has long been considered the paradigm for Monte Carlo 
simulations in statistical physics, yet 
this method faces some severe drawbacks in situations where the sequence of states created by the
Markovian chain leads to very repetitive dynamics, 
i.e., dramatically low acceptance rates
and exponentially diverging autocorrelation times: this makes it necessary
to simulate systems over exceedingly long runs in order to obtain good statistics and reliable estimates of 
thermodynamical averages (see, for example, \cite{LandauBinderBook2000} and the contribution by
Krauth in \cite{Flyvjerg1996} for an introductory review). 
This is the case when one comes to simulating
systems with rugged free energy landscapes, e.g., polymers, proteins, and disordered systems including spin-glasses, 
for the dynamics may then get trapped in one of numerous local minima, especially at low temperature. 
 One experiences
similar behavior  when simulating first-order phase transitions
(the so-called supercritical slowing down \cite{BergNeuhaus1992}), 
where the tunneling time between coexisting
phases grows exponentially with the system size, due to the increasingly high free energy barrier to be 
overcome (e.g., \cite{Binder1982}).   

Since slow dynamics mainly results from the combination of weighting the Markovian chain with 
Boltzmann weights and using local updates, there have been several attempts to devise
 efficient update algorithms based on \textit{global updates}, e.g., cluster algorithms, which in the
 case of continuous transition decrease critical slowing down by 
several orders of magnitude (see \cite{SwendsenWang1987,Wolff1989}; also a LR implementation in \cite{LuijtenBlote1995}).
On the contrary, multicanonical methods \cite{BergNeuhaus1991,BergNeuhaus1992,Berg1992,Janke1998} 
are based on random walks in the energy landscape, irrespective of the particular move update utilized, 
whereby a \textit{flat} energy distribution is now sampled. 
First, this results in the algorithm quickly sampling a much wider phase space than in the canonical case, 
 by allowing the system to cross any free energy barrier.
Second, this allows the density of states to be computed over the whole energy axis, thus
extending the reliability of reweighting procedures over a much wider range of temperature
than in the case of standard histogram methods \cite{FerrenbergSwendsen1988}, where poor
histogram sampling at low-energy usually induces strong statistical bias.
As opposed to multihistogramming \cite{FerrenbergSwendsen1989}, a single run is needed to cover 
the energy range of interest.
Once a reliable estimate
of the density of states has been obtained, it is then straightforward to compute thermodynamical
functions otherwise hardly within reach of canonical simulations, e.g., canonical entropy and free energy.
It is noteworthy that this simulation technique actually belongs to a larger class of algorithms 
 called \textit{generalized-ensemble} algorithms, which encompasses 
variants based on random walk in the entropic variable 
("$1/k$ ensemble" or "entropic sampling" algorithms \cite{Lee1993,HesselboStinchcombe1995}),
 or the temperature variable (e.g., "simulated tempering" \cite{Marinari1992,LyubartsevEtAl1992}).
  
\subsection{Rationale}
  
The rationale behind the multicanonical algorithm is the generation of a Markovian chain of states 
$\{\sigma_i\}$, whose weights $W_{mu}(E(\sigma_i))$ are tweaked so that one eventually gets a 
flat energy histogram, i.e. if $P(E)$ denotes the probability in energy and 
$n(E)$ is the density of states,
$$
	P_{mu}(E) \propto n(E) W_{mu}(E) = \mbox{const.}
$$
Since $n(E)$ usually increases drastically with energy, low-energy states are thus 
sampled much more often than high-energy ones.

Following Berg in \cite{Berg1996}, we compute $W_{mu}(E)$ through an iterative procedure, starting from an initial canonical
simulation at inverse temperature $\beta_0$. $\beta_0$ indirectly sets the energy 
below which the energy histogram is to be flat, i.e. $E_{max} = \meanval{E}_{\beta_0}$.
Thus, $kT_0=1/\beta_0$ must be chosen high enough
 to ensure that the final energy histogram spans a suitably large energy range upward, 
 e.g., reaches the energy of the disordered phase in the case of a first-order transition,
 and extends even further away if one wants to observe with satisfying accuracy 
 the free energy plateaus signaling the limit of metastability. 
For convenience, we now define an effective Hamiltonian $H_{mu}(E)$, so that 
$$
	W_{mu}(E,\beta_0) = e^{-\beta_0 H_{mu}(E)}.
$$
Hence, multicanonical simulation can be envisioned as a canonical simulation at inverse temperature
$\beta_0$ with the usual Boltzmann weight, provided the original Hamiltonian is replaced by an effective
Hamiltonian to be determined iteratively. As a side note, a cluster implementation in the framework of the multicanonical algorithm
is thus far less straightforward,
since this effective Hamiltonian has fundamentally a global nature, whereas canonical simulations explicitly preserve the locality
of the original Hamiltonian (see, e.g., the multibond approach in \cite{JankeKappler1995,Janke1998}). 

Denoting $H^{\infty}_{mu}(E)$ as the true estimate of the effective Hamiltonian, we
may thus write
$$
	n(E) \propto e^{\beta_0 H^{\infty}_{mu}(E)}.
$$
The microcanonical inverse temperature $\beta(E)$ may be easily related to $H^{\infty}_{mu}(E)$, 
as we have (assuming $k=1$) 
$$
	\beta(E)=\frac{d\ln n(E)}{dE}=\beta_0 \frac{d H^{\infty}_{mu}(E)}{dE}
$$
Since the dynamics of the Markovian chain is governed by the transition rate
$W(a\rightarrow b)=\min(1, \exp\{\beta_0 [H_{mu}(E_a)-H_{mu}(E_b)]\})$, we may
write, for two states infinitely close in energy, i.e., whenever $E_b=E_a+\delta E$, 
$W(a\rightarrow b)=\min(1, \exp[-\beta(E_a)\delta E])$.
Hence it is the {\em microcanonical temperature} which is the relevant quantity where the
dynamics (e.g., the acceptance rate) of the multicanonical algorithm is concerned. 

\subsection{Iteration scheme}

We initially set $H_{mu}^0(E)=E$, or equivalently $\beta^0(E)=\beta_0$, 
as this indeed corresponds to a 
canonical simulation at temperature $1/\beta_0$. 
At step $i$, a simulation is performed using a 
Boltzmann weight with effective Hamiltonian $H^i_{mu}(E)$; then an energy histogram $N^i(E)$ is 
eventually computed using independent samples. Incidentally, taking truly independent samples proves 
useful during the late stages of the iteration scheme only, where the aim is then to refine a nearly flat histogram. 
During early iteration steps, histograms may be computed using nonindependent samples without significantly
 affecting the convergence. 
We now denote $E^i_{min}$ as the lowest energy level that was reached throughout 
the previous runs, including step $i$: this is the energy level below 
which $H_{mu}^{i+1}(E)$ will have to be predicted,
since no histogram data are available inside this energy range. Issues regarding adequate predictor
choice will be considered later on in this section. 
The rules for updating $H_{mu}^{i+1}$ at step $i+1$ from $H_{mu}^{i}$ at step $i$ 
are based on the following equations. For $E \ge E_{max}$, $H_{mu}^{i+1}(E) = E$, i.e., the
dynamics is canonical-like at inverse temperature $\beta_0$ for all iteration steps. 
For $ E^i_{min} \le E < E_{max}$, 
\begin{equation}
	\label{equ:update_beta_g0cum}
	\beta^{i+1}(E) = \beta^{i}(E) + \frac{\hat g_0^i}{\delta E} \ln \frac{N^i(E+\delta E)}{N^i(E)}, 
\end{equation}
where  
$$
	\hat g_0^i = \frac{g_0^i}{\sum_{k=1}^i g_0^k}
$$
and $g_0^k$ is a \textit{raw} inverse damping factor proportional to the reliability of the $k$th histogram.
It has been shown in \cite{Berg1996}, following an error calculation argument, that 
$$
	g_0 = \frac{N(E) N(E+\delta E)}{N(E) + N(E+\delta E)}
$$
provides an estimator proportional to the inverse of the variance of
$\beta^{i+1}(E)$. 
Once $\beta^{i+1}(E)$ is known, $H^{i+1}_{mu}(E)$ is derived by a mere integration starting
from the initial condition $H_{mu}(E_{max})=E_{max}$.
Finally, for $E < E^i_{min}$, $H_{mu}^{i+1}(E)$ will have to be computed using a
suitably chosen predictor, until at last $E^i_{min}$ becomes equal to the ground state energy.
A cubic spline is then fitted to $H_{mu}(E)$ at every bin center, and
this curve is used to compute acceptance probabilities during the next run.

 It can be seen that Eq.~(\ref{equ:update_beta_g0cum}) leads to a
steady state whenever $N(E)$ is constant over the energy range of interest.
Writing a recursion equation involving $\beta(E)$ instead of $H_{mu}(E)$, together
with the inclusion of a damping factor, allows one to handle the situation 
where some bins have null entries, a case which otherwise leads to a fairly spiky graph
for $H_{mu}(E)$ and inconsistent dynamics. 
\textit{Accidental} null entries at energy values $E$ or $E+\delta E$ will simply leave
$\beta(E)$ unchanged, and the corresponding parts of $H_{mu}(E)$ thus move as a block.
Since acceptance rates hinge on the microcanonical temperature, this in effect 
drastically reduces bias on the dynamics.
Considering a small set of histogram bins that are copiously filled for the first time during a given iteration run  
(e.g., high-energy bins during the early iteration runs whenever we begin
with a canonical simulation), we see that the related \textit{cumulative} inverse damping factor first soars and 
produces a great amount of change in $\beta(E)$ in the couple of runs that follow, and then decays progressively to zero
as these bins continue to be filled.
By taking into account all the data that have been sampled up to step $i$, this 
modified recursion both clearly stabilizes the algorithm and reduces relative errors due
to poor histogram sampling.    

Choosing the most appropriate value of the histogram bin width results 
 from a trade-off between resolution and computation time. 
A higher resolution on the one
hand guarantees good histogram flatness, and is especially crucial at low energy levels,
where the density of states displays a rugged graph. 
On the other hand, we impose a fixed number of independent samples per histogram bin,
 so as to give the histogram variance an acceptably low value;
hence a low $\delta E$ implies more simulation steps per iteration. Our approach is thus to
first choose a fairly high $\delta E$, e.g., one yielding around $20$ bins, during the early stages of the iteration process in order to
obtain a rough picture of the density of states, and then to progressively reduce $\delta E$ once the
ground state has been reached. As will become obvious in Sec.~\ref{sec:free_energy}, 
the ultimate value of $\delta E$ deeply affects the attainable precision on the computation of spinodal points,
since the latter is based on a precise location of free energy plateaus, and this indeed
 entails having enough bins belonging to a given plateau. 
As a rule of thumb, the best compromise is then to obtain between $100$ and $300$ histogram bins in the final stage,
with the number of bins increasing as the $\sigma$ value corresponding to the second-order regime is approached.
\begin{figure}
	\centering
	\InsertFig{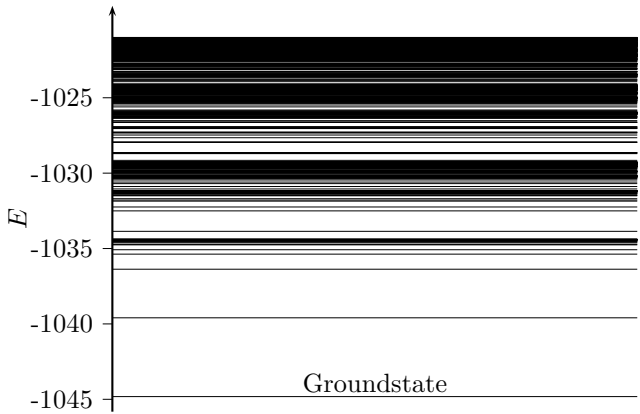}
	\caption{Lowest energy levels for $q=5,\sigma=0.5,N=400$, computed by sorting
	energy samples from a long simulation run. Each level is drawn as a horizontal line.}
	\label{fig:first-energy-levels-q5s05N400}
\end{figure}

In this view, the unequal spacing of energy levels in LR spin models deserves specific attention. 
As witnessed in Fig.~\ref{fig:first-energy-levels-q5s05N400}, large energy gaps separate
isolated energy levels or tiny groups thereof in the vicinity of the ground state, 
whereas the distribution gradually turns into a near continuum above
$E \sim -1025$. Setting a low $\delta E$ value leads in turn to
\textit{nonaccidental} null entries in those bins located inside energy gaps, whereby $\beta(E)$ never
gets updated at isolated energy levels and $g_0$ is always zero. Since the graph of the density of states looks indeed fairly
wrinkled near the ground state, and the dynamics there is noticeably sensitive to even the smallest departure 
of $H_{mu}(E)$ from the ideal line, we would then observe a sharp steady peak in the lowest part of the energy histogram,
which the present recursion would not be able to suppress. One could trivially think of working this out 
by implementing variable-width bins that would span energy gaps. This is, however, impracticable
 since the distribution of 
energy levels is not known prior to starting the iteration process (for this is precisely what we intend to compute with
the density of states). 
To circumvent this limitation, we have modified the previous recursion so that null entries are always skipped, however
accidental or nonaccidental they may be. Denoting by $E_a$ and $E_b$, with $E_a<E_b$, the centers of 
histogram bins located on each side of a set of contiguous empty bins, we have
\begin{equation}
	\label{equ:update_beta_g0cum_EaEb}
	\beta^{i+1}(E_a) = \beta^{i}(E_a) + \frac{\hat g_0^i(E_a)}{E_b-E_a} \ln \frac{N^i(E_b)}{N^i(E_a)},
\end{equation}
where $\beta(E_a)=\beta_0 \{H_{mu}(E_b)-H_{mu}(E_a)\}$ and we now impose 
$$
	g_0(E_a) = \frac{N(E_a) N(E_b)}{N(E_a) + N(E_b)};
$$
hence $g_0$ can never be zero.
In order to avoid losing details of the shape of $H_{mu}(E)$ for $E_a<E<E_b$ that were
possibly collected during previous runs, we update $H_{mu}(E)$ through a linear difference scheme,
$$
	\delta H_{mu}(E) = \frac{\delta H_{mu}(E_b)-\delta H_{mu}(E_a)}{E_b-E_a} (E-E_a) + \delta H_{mu}(E_a),
$$
where $\delta H_{mu}(E) = H_{mu}^{i+1}(E) - H_{mu}^{i}(E)$. While this has obviously no effect where nonaccidental null entries
are concerned, this favors quicker convergence during the early runs where the inadequate shape of $H_{mu}(E)$ is more likely
to produce empty bins. 

The iteration process stops whenever the energy histogram has become suitably flat over the energy range of interest, namely,
between the ground state energy and $E_{max}$ for our purpose. We evaluate this property by computing the standard 
deviation of histogram entries, as well as the same quantity for the logarithm of histogram entries restricted to
nonempty bins. The latter seems to be a better indicator since it is sensitive to both poorly populated bins and
histogram peaks, whereas the former increases only with rather spiky histograms. In addition, we estimate the degree of 
convergence of the algorithm by computing the mean square distance between $H_{mu}^i(E)$ and $H_{mu}^{i+1}(E)$ after
the ground state has been reached.  
We then compute a threshold value for each indicator by trial and error, based on a couple of short runs
for various lattice sizes and bin widths.   

\subsection{Reweighting procedure}

Once $H_{mu}(E)$ has been satisfactorily computed,
a long production run is performed using this effective Hamiltonian in place of the original
one,and then estimates of thermodynamical quantities of interest at inverse temperature $\beta$
are computed using a reweighting scheme,
i.e., formally,
$$
	\meanval{A}_\beta = \frac{\sum_E \meanval{A}_E n(E) e^{-\beta E}}{Z(\beta)},
$$
where $\meanval{A}_E$ denotes the microcanonical average of $A$ at energy $E$,
and the partition function is given by $Z = \sum_E n(E) e^{-\beta E}$. 
The best estimate for the density of states $n(E)$ is provided by $n(E) \propto N(E) e^{\beta_0 H_{mu}(E)}$,
where $N(E)$ stands for the number of bin entries at energy $E$ computed from the production run. In order to avoid numerical
overflows, as well as to suppress bias resulting from possibly strong variance on microcanonical averages, 
we found it more appropriate to compute $\meanval{A}_\beta$ from a sum running over samples instead of
energy bins, i.e., $\meanval{A}_\beta = \sum_i A_i  w(E_i) / \sum_i w(E_i)$,
where $w(E_i)=e^{\beta_0 H_{mu}(E_i)-\beta E_i-K}$.
$K$ is then determined so as to avoid both numerator and denominator overflows. Providing that the histogram
sampled during the production run is flat to a good approximation, the maximum in $e^{\beta_0 H_{mu}(E)-\beta E}$ is
reached whenever $dH_{mu}(E)/dE \sim \beta/\beta_0$,
which yields the energy value at which $K$ is to be computed. In addition, since the reweighting scheme
involves an exponential contribution of $H_{mu}(E)$, the resulting curve $e^{\beta_0 H_{mu}(E)-\beta E}$
is strongly peaked around the maximum; hence it is clear that only histogram points
in the vicinity of this maximum contribute to $\meanval{A}_\beta$. In effect, we found 
that the existence of two distinct
maxima, or equivalently of two energy values for which $\beta(E)$ has the same value, coincides
with the occurrence of a first-order phase transition. 

Following the same reweighting procedure, we compute partial free energy functions, 
i.e., $F(\beta, m)$ where $m$ is the order parameter, and reweighted histograms of the 
energy, i.e., $N_{rw}(\beta,E)$.
The partial partition function is straightforwardly
derived from a partial sum over samples having the prescribed order parameter,
\begin{equation}
	Z(\beta,m) =  \sum_i e^{\beta_0 H_{mu}(E_i)-\beta E_i} \delta_{m,m_i},
	\label{equ:free_energy_mag_reweighting}
\end{equation}
which then yields $F(\beta,m) = -\ln Z(\beta,m) / \beta$. Similarly,
a reweighted histogram of the energy is obtained from  
$N_{rw}(\beta,E) =  N(E)  e^{\beta_0 H_{mu}(E)-\beta E}$.

\subsection{Predictor choice}

We now discuss some issues related to the choice of an efficient predictor for $E < E_{min}$.
For small lattice sizes, we initially feed the algorithm with an effective Hamiltonian $H_{mu}(E)=E$, 
and the objective is then to find an appropriate trade-off between
 speeding up the convergence of $E^i_{min}$ toward the ground state and avoiding algorithm instability. 
 While the former demands that $H^i_{mu}(E)$ have a sufficiently high slope below $E^i_{min}$,
 the latter still requires that the algorithm remain ergodic to a suitable extent.
Our implementation relies on a  
first-order predictor, $H_{mu}(E)=a + b E$, 
and we impose continuity on $H_{mu}(E)$ at $E_{min}$. 
 The simplest approach is then to choose 
a  predictor slope so that continuity on $H'_{mu}(E)$ is enforced at
$E=E_{min}$, i.e., $b = \beta(E_{min})/\beta_0 $.
While $E_{min}$ reaches the ground state rather quickly using this predictor, 
 the dynamics often gets locked in very low
energy levels due to the particularly steep slope of $H_{mu}(E)$ in the vicinity of
 the ground state. The time needed by the iteration scheme to recover from this deadlock and obtain
 a flat histogram thus becomes prohibitive. 
On the other hand, choosing $b=1$ leads to the smoothest yet slowest
convergence, and avoids deadlock issues. An efficient compromise is thus to ensure a "weak" continuity at
$E_{min}$, i.e., by computing the slope of the predictor using a least-square
scheme based e.g., on the first 10\% of points above $E_{min}$. 

For large lattice sizes where reaching the ground state energy can become time consuming, 
we resort to a "scaling trick" whereby $H_{mu}(E)$ is initially guessed from
the density of states obtained at a smaller lattice size. 
This approach was initially mentioned by Berg
and Neuhaus \cite{BergNeuhaus1992}, and claimed to work perfectly within the framework of a study of
the two-dimensional ten-state Potts model with nearest-neighbor
interactions where the energy is additive to a perfect extent. 
The presence of LR interactions, however, slightly worsens the case, especially at low $\sigma$.  
The scaled density of states is computed as follows.
Let us consider, for the sake of
simplicity, two systems $\Sigma$ and $\bar\Sigma$ with respective lattice sizes 
$L=N$ and $\bar L=2N$, and let us divide
the latter into two subsystems $\Sigma_1$ and $\Sigma_2$ of equal size $L$. 
Since $H_{mu}(E)=kT_0 \ln n(E)$, where $n(E)$ stands for the density of states, we
 have to compute $\bar n(E)$ for $\bar\Sigma$ as a function of $n(E)$ for $\Sigma$.
Neglecting the interaction between subsystems $\Sigma_1$ and $\Sigma_2$, and denoting by $E_1$ 
 the energy of $\Sigma_1$, the density of states for $\bar\Sigma$ just reads
$ \bar n(E) \simeq \sum_{E_1} n(E_1) n(E-E_1)$, which yields 
\begin{align*}
	\beta_0 \bar H_{mu}(E) &\simeq \ln \sum_{E_1} e^{\beta_0 [H_{mu}(E_1)+H_{mu}(E-E_1)]}\\
		& \sim \ln \frac{1}{\delta E}\int dE_1 e^{\beta_0 [H_{mu}(E_1)+H_{mu}(E-E_1)]},
\end{align*}
where $\delta E$ is the energy histogram bin width.
Providing that $n(E)$ is a monotonic and rapidly increasing function of $E$, we may use a
saddle-point approximation to evaluate the former sum. The maximum of  
$H_{mu}(E_1)+H_{mu}(E-E_1)$ is reached whenever $E_1=E/2$; hence
\begin{equation}  
	 \bar H_{mu}(E) \simeq   2 H_{mu}\left(\frac{E}{2}\right)   
	 + kT_0 \ln \frac{\sqrt{\pi/|H_{mu}''\left(E/2\right)|}}{\delta E}
	 \label{equ:scaling-trick}
\end{equation}
\begin{figure}[bt]
	\centering
	\InsertFig{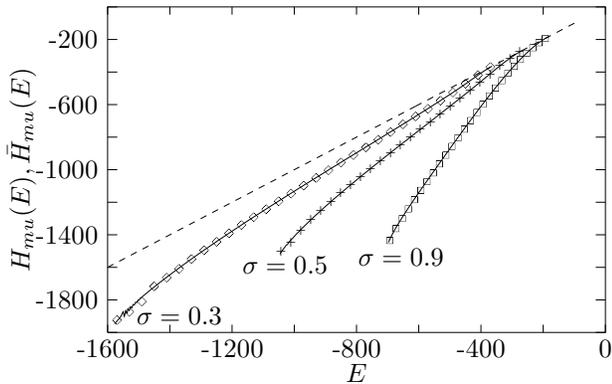}
	\caption{Dots indicate the initial guesses $\bar H_{mu}(E)$ that were 
	fed into the iteration scheme at $L=400$, $q=5$, and $\sigma=0.3(\Diamond),0.5(+),0.9(\square)$. 
	Each initial guess was computed using Eq.~(\ref{equ:scaling-trick}), i.e., by scaling a true estimate obtained at $L=200$.
	Solid lines show true estimates $H_{mu}(E)$ as obtained after the whole iteration scheme at $L=400$ converged.
	The straight dashed line sketches the original Hamiltonian, i.e., $H_{mu}(E)=E$.}
	\label{fig:scaling-predictor}
\end{figure}
\begin{figure}[bt]
	\centering
	\InsertFig{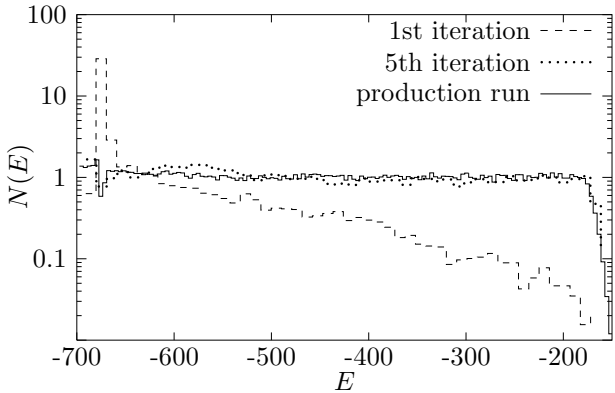}
	\caption{Energy histogram as computed after indicated runs, for
	$q=5,\sigma=0.9,L=400$ spins, using
	Eq.~(\ref{equ:scaling-trick}) to compute the initial effective Hamiltonian $\bar H_{mu}(E)$ from a previous run at
	$L=200$ spins. Labeling on $y$ axis indicates normalized probabilities.}
	\label{fig:flat_histo}
\end{figure}
This expression may be readily extended to lattice sizes that are any multiple of the original size.
Figure~\ref{fig:scaling-predictor} sketches results obtained for $q=5$ and $\sigma=0.3,0.5,$ and $0.9$. 
A series of iteration runs was first conducted with $L=200$ spins in 
order to obtain an estimate of $H_{mu}(E)$ for this lattice size, then this estimate
was scaled using Eq.~(\ref{equ:scaling-trick}) and used as the initial guess $\bar H_{mu}(E)$ 
for the next series of iteration runs at $L=400$. Equation~(\ref{equ:scaling-trick}) yields a very acceptable guess for $\sigma=0.9$, 
and the two curves are hardly distinguishable from each other.
As illustrated in Fig.~\ref{fig:flat_histo}, the energy histogram becomes nearly flat within five iterations. 
For $\sigma=0.3$ and $0.5$, the agreement remains quite satisfying, yet 
the initial guess falls slightly below the true estimate at low energy levels, and the lowest-energy bins
are exceedingly enhanced during the first iteration runs. More iteration runs are thus required to obtain a perfectly flat
histogram as $\sigma$ is decreased, and the benefit of this approach in effect becomes negligible for $\sigma\leq 0.3$.
Indeed, the algorithm then spends a great number of iteration steps being trapped in low energy levels, seeking
to rectify the shape of the density of states in this energy region until convergence is obtained:
starting from an initial canonical effective Hamiltonian actually yields better performances.
Since, for systems with LR interactions, computation time scales with $L^2$, using this "scaling trick" thus
greatly reduces the time needed for proper convergence, at least for $\sigma> 0.3$, and partially compensates for the lack
of a hybrid multicanonical-cluster algorithm dedicated to LR models.

\subsection{Algorithm performance}

In order to measure the performance of our implementation,
we have computed a dynamical exponent $z$ defined as the scaling exponent of
a relevant characteristic time $\tau$ of the simulation, i.e., $\tau \propto L^z$, where $L$ denotes the lattice size:
while for second-order transitions it is widely known that the integrated autocorrelation time
 represents such a relevant time, for first-order transitions the tunneling time through
 the energy barrier ($\tau_{tun}$) proves to be a more meaningful indicator \cite{JankeKappler1995}. We
 define the latter as one half of the average number of Monte Carlo steps per spin (MCS) 
 needed to travel from one peak of the reweighted energy
histogram ($N_{rw}(\beta,E)$) to the other, with the temperature being set to the transition temperature. 
 Tunneling time is expected to grow exponentially with $L$ for canonical
 algorithms, and to scale as a power law of $L$ for multicanonical algorithms \cite{BergNeuhaus1992}.
In both cases, it appears that the chosen characteristic time is a good indicator of
how quickly the demands in CPU time should grow with increasing lattice size: for second-order transitions,
this is the time needed to generate truly independent samples, 
while for first-order transitions, this tells us at what rate the dynamics spreads out
over the energy barrier and thus to what extent samples get efficiently picked from the two phases in coexistence. 

The integrated autocorrelation time 
is computed by using the well-known time-displaced correlation function 
which displays an exponential-like
short-time behavior, namely, $\Phi_{mm}(t) \sim e^{-t/\tau}$; 
 $\tau$ is then derived from a simple integration scheme.
Since the latter function makes sense within equilibrium only, we first discard $n$ thermalization steps, where
$n$ is obtained by using
 the nonlinear relaxation function that describes the approach to equilibrium \cite{LandauBinderBook2000}
 and averaging over several dry runs.
An interesting point regarding multicanonical simulations is that, since they are random walks in energy space, 
 "thermalization" (although this term is no longer appropriate as far as generalized ensemble algorithms are concerned)
 always occurs rather rapidly; 
 simulations based on a nearly flat histogram have shown that a value of $1000$ MCS is indeed appropriate on average.

\begin{figure}[bt]
	\centering
	\InsertFig{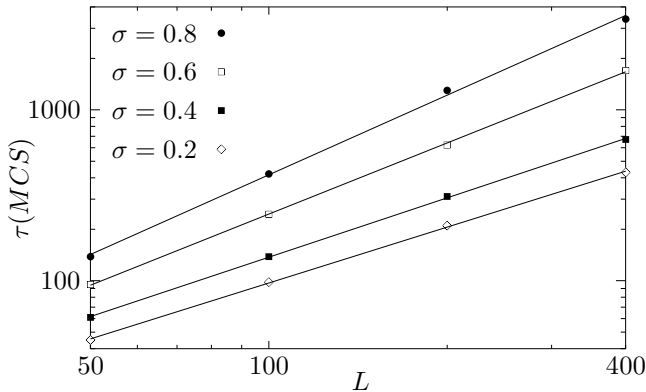}
	\caption{Integrated autocorrelation time $\tau$ vs lattice size $L$ for $q=7$
		and $\sigma=0.2,0.4,0.6,0.8$.
		Dynamic exponents computed from a fit to $L^z$ are
		$z=1.09(1), 1.15(1), 1.38(1), 1.55(1)$, respectively.
		}\label{fig:autocorrel_times}
\end{figure}

\begin{figure}[bt] 
	\centering
	\InsertFig{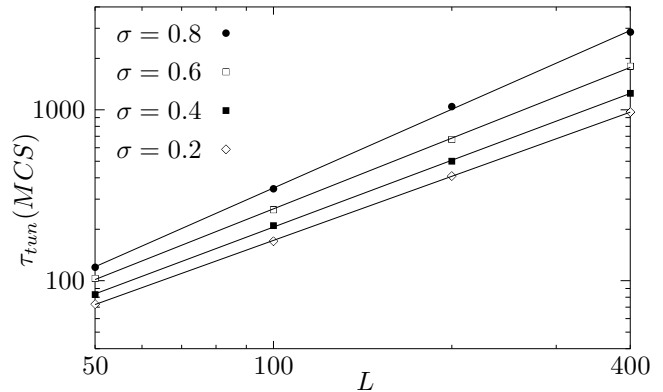}
	\caption{Tunneling time $\tau_{tun}$ vs lattice size $L$ for $q=7$ and 
		$\sigma=0.2,0.4,0.6,0.8$.
		Dynamic exponents computed from a fit to $L^z$ are 
		$z=1.25(1),1.30(2),1.37(1),1.53(1)$, respectively. 
		}\label{fig:tun_time}
\end{figure}

Results for $q=7$ and $\sigma$ lying between $0.2$ and $0.8$ are shown in Fig.~\ref{fig:autocorrel_times} for
integrated autocorrelation times, and in Fig.~\ref{fig:tun_time} for tunneling times.
The slight dispersion in the power-law fits
arises from the fact that simulations at larger sizes were conducted with
a higher number of MCS between measurements in order to reduce memory overhead.
Where computing tunneling times is concerned, 
this results in some tunneling events being possibly skipped and
the average tunneling time being overestimated.
Both figures show, however, that a power-law behavior is perfectly plausible.
In the case of first-order transitions, the reduction in simulation costs is thus drastic in comparison
with standard canonical algorithms.  

For both indicators, we obtain an average $z$ slightly above $1.0$ for $\sigma=0.2$, 
yet $z$ increases smoothly with decreasing range of interaction.
This may be accounted for by the fact that spatial and time correlations grow as we depart 
from the MF regime and approach the SR one. As for tunneling times, the prefactor turns out
to be slightly higher near the MF regime, and $z$ increases at a lower rate with increasing $\sigma$
than is the case for autocorrelation times.   

Since there are no other numerical studies of LR models based on multicanonical algorithms to our knowledge, 
comparison is limited to
 estimates obtained for SR models. For the three-state Potts model,  
canonical simulations using local updates
led to $z=2.7$ \cite{Williams1985}, while 
Swendsen and Wang obtained $z \sim 0.6$ using their percolation cluster algorithm \cite{SwendsenWang1987}. 
For further comparison, the Metropolis algorithm applied to a SR Ising chain in $d=2$ and $d=3$ yielded a value of 
$z$  slightly above $2$ \cite{WanslebenLandau1991}, whereas Wolff's cluster
algorithm led to $z \sim 0.27$ \cite{Wolff1989a}.
While our value is slightly greater than in the case of cluster implementations, it is worth underlining that our
multicanonical implementation yields reliable statistics within a single MC sweep, whereas several are needed in the case
of a standard canonical simulation, whatever reweighting procedure may be used.

\section{Numerical results} \label{sec:results} 

We have conducted multicanonical simulations for $q \in [3, 9]$, using for each value
 of $q$ an appropriate set of $\sigma$ parameters between $0.3$ and $0.9$, 
 so that we could observe strong and weak first-order transitions, as well as continuous ones.
 For $q=3$, some simulations were performed with $\sigma>1.0$ in order to investigate the crossover from LR to SR regimes.
Once the density of states had been determined using the iteration process described above,
a production run was performed for lattice sizes between $L=50$ and $L=400$. The number of MC sweeps needed
for each production run was computed so as to yield approximately $5\times 10^{4}$ truly independent samples.
In this view, rapidly increasing autocorrelation times in effect
restricted our work to lattice sizes $L\leq 400$.

\subsection{Free energy functions and FSS } \label{sec:free_energy} 

As already stated in the Introduction, a precise determination of the tricritical value $\sigma_c(q)$ is a real challenge, 
due to the weakening of the first-order transition as $\sigma_c$ is approached from below. 
This makes traditional indicators e.g., latent heats or energy jumps, fairly
inefficient, since observing clear jumps in the vicinity of the tricritical value entails simulating huge lattices.  
Glumac and Uzelac in \cite{GlumacUzelac1997_1998} used three less traditional estimators, namely, the interface free energy,
the specific heat, and the reduced fourth-order Binder cumulant, which all turned out to be less sensitive to this weakening: in particular,
the last quantity defined as $U_L = \meanval{E^4}/\meanval{E^2}$ is expected to reach a nontrivial constant 
$U_\infty \neq 1$ as $L \rightarrow \infty$ at a first-order transition only \cite{ChallaLandauBinder1986}; by extrapolating
to the thermodynamic limit from measures taken at different lattice sizes, 
they found $\sigma_c$ to fall between $0.6$ and $0.7$ for
 the three-state model. 
 Still and all, this approach imposes simulating fairly large lattices (around $L=3000$) for the
 extrapolation procedure to be reliable, let alone the fact that Binder cumulants may
 experience uncontrollable crossover effects \cite{LeeKosterlitz1990}.
Due to the modest lattice sizes that are within reach of our local update algorithm, we rather  
resort to an approach based on the location of spinodal points, which may be accurately determined
already for medium lattice sizes.
In marked contrast to multihistogram techniques, the multicanonical method indeed allows one to
 obtain partial free energy functions (or, equally, reweighted histograms of the energy) over a range of temperature 
 which extends fairly far away from the transition temperature, with remarkably modest numerical effort.

 The basis of our method relies on the fact that the temperature difference between the two spinodal points will tend to zero as $\sigma_c$
 is approached, since there are no metastable states in the case of continuous transitions. Stated differently, the conditions under which
metastability occurs, i.e., both first and second derivatives of the partial free energy are zero, are met only at 
the critical point for a continuous transition: hence metastable states merge into a single large minimum as the 
first-order transition turns into a second-order one. Such behavior has indeed been widely observed,
e.g., for liquid-vapor transitions near the critical point, and is borne out by our MF calculation. 

For a given set of $(q,\sigma)$ parameters, we determine the location of the spinodal points by first computing 
a partial free energy function of the order parameter [$F(kT,m)$, see Eq.~\ref{equ:free_energy_mag_reweighting}] 
 over a large temperature range.
Alternatively, we make use of a similar function of the energy, 
i.e., $F_e(kT,E)=-\ln N_{rw}(kT,E)$, where $N_{rw}(kT,E)$ denotes the reweighted
histogram of the energy. 
While the latter function plays the 
same role as the partial free energy of the magnetization, it yields a higher
precision at low $q$, as we will witness in a moment.
The limit of metastability at finite lattice size is then defined by $dF_e/dE=d^2F_e/dE^2=0$, or alternatively $dF/dm=d^2F/dm^2=0$: 
for a first-order transition, this condition
is met at two temperatures $T_1$ and $T_2$ which satisfy the inequality $T_1 < T_c < T_2$, where $T_c$ denotes the transition temperature.

\begin{figure}[bt]
	\centering
	\InsertFig{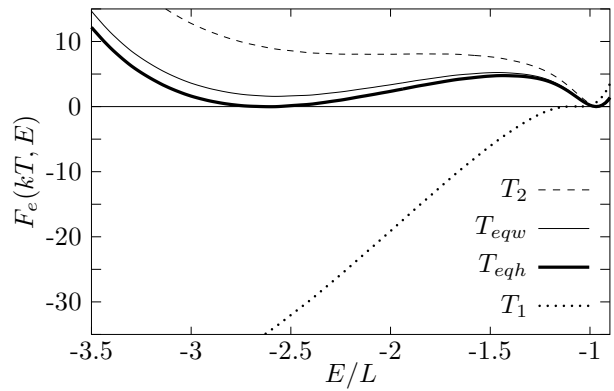}
	\caption{Graphs of $F_e(kT,E)=-\ln N_{rw}(kT,E)$ for $q=5,\sigma=0.3, N=400$,
	at four characteristic temperatures:
	$T_1, T_2, T_{eqh}$, and $T_{eqw}=T_c$ denote the temperatures of the two metastable states,
	and the temperature of equal peaks heights, and that of equal peak weights, respectively.
	$E/L$ denotes the energy per spin.}
	\label{fig:rew_histo_q5s4N400}
\end{figure}

Since these free energy functions usually have rather rugged graphs, we first
filter out rapid oscillations by means of a linear smoothing filter, whose
order is computed so that we are left
with at most three extrema over the whole temperature range of interest. 
By continuously varying $kT$ within this range, we determine the temperature of each metastable state by  monitoring
the change in the number of minima (see Fig.~\ref{fig:rew_histo_q5s4N400}). 
In contrast to \cite{GlumacUzelac1997_1998}, the transition temperature $T_c(L)$ is then obtained 
by imposing that the number of bin entries in $N_{rw}(E)$ 
be the same below and above the energy corresponding to the maximum of $F_e(kT,E)$. This corresponds to the
so-called equal-weights condition as proposed by Lee and Kosterlitz in \cite{LeeKosterlitz1990}, 
and is equivalent to the condition that the average energy be the arithmetic mean of 
the energy of each phase.
For the sake of comparison, however, we also compute the temperature $T_{eqh}(L)$ 
at which both minima of $F_e(kT,E)$ have the same
value. We then proceed with the computation of similar quantities
using $F(kT,m)$, and we estimate statistical errors using a bootstrap procedure.

\begin{figure}[bt]
	\centering
	\InsertFig{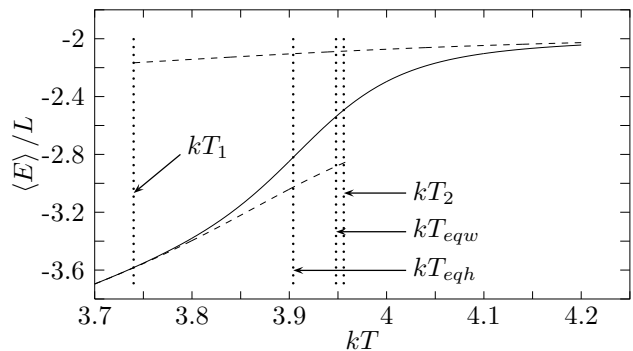}
	\caption{Average energy per spin for $q=3,\sigma=0.2,L=400$, 
	 computed over both phases (solid line), 
	 ordered phase only (lower dashed line), 
	 and disordered phase only (upper dashed line).
	 Vertical dotted lines indicate the four characteristic temperatures: from left to right,
	 lower limit of metastability ($kT_1$), 
	 transition temperatures (equal heights $kT_{eqh}$, then equal weights $kT_{eqw}$), 
	 and upper limit of metastability ($kT_2$).
	 }\label{fig:mean_ene_per_phase}
\end{figure}

Graphs of the free energy $F_e(kT,E)$ in Fig.~\ref{fig:rew_histo_q5s4N400} show that the peak and the plateau corresponding to the disordered phase
are much narrower than those of the ordered phase. As a result, the precision in the determination of the temperature $T_1$ of the lowest
metastable state is fairly lower than that of the upper metastable state ($T_2$). This asymmetry increases with increasing $q$, and in effect
precludes the use of reweighted histograms for the estimation of spinodal points at $q>7$. For $q=9$, we thus relied 
on the extrema of the partial free energy $F(kT,m)$, since this function then becomes nearly symmetric and displays
peaks that are well separated.
Incidentally, the asymmetric shape of $F_e(kT,E)$ can be accounted for by the fact that specific heats have a different magnitude 
in the disordered and ordered phases, since this thermodynamic quantity is simply proportional to the standard deviation
of the associated Gaussian peak \cite{ChallaLandauBinder1986}. 
This may be readily observed by reweighting thermodynamical averages
over a single phase at a time, once the maximum of $F_e(kT,E)$ which separates the two phases has
been located.
Figure~\ref{fig:mean_ene_per_phase} shows how this procedure was applied to the computation of the mean energy per spin of each subphase
for $q=3, \sigma=0.2$, and $L=400$ spins. A simple visual inspection allows one to assess a much lower specific heat for the disordered phase
than for the ordered phase.  

At finite lattice size, however, all these temperatures experience a distinct 
shift proportional to the distance from the thermodynamic limit. Assuming that
 the FSS theory developed in \cite{ChallaLandauBinder1986} for first-order transitions is also valid in the LR case, 
 we therefore compute temperatures at infinite lattice size by assuming power-law corrections in $1/L$. 
We also expect temperatures defining the limit of metastability to obey the same scaling behavior, 
although the phenomenological theory proposed in \cite{ChallaLandauBinder1986} 
does not explicitly handle them.  
 The inclusion of a second-order term proves necessary in order to to obtain satisfying fits, due to 
the presence of small lattice sizes in our set of data.  
Yet, interestingly enough, fitting finite-size temperatures to a power law of the form $T(L)=T(\infty)+a L^b$
yields very similar extrapolated values, with discrepancies smaller than $0.1\%$, i.e., within
our range of uncertainty. In addition, we observed that $F_e(kT,E)$ and $F(kT,m)$ led to distinct finite-size shifts,
with the latter function easily allowing one to drop second-order correction terms without much affecting the final result.

\subsection{Transition temperatures} \label{sec:temperatures_first_order} 

For the sake of completeness, we also compute transition temperatures by relying on two other estimators,
namely, the magnetic susceptibility, which for magnetic systems has more pronounced peaks than the specific heat,
and Binder cumulants of the magnetization defined as $U^{(4)}=1 -\meanval{m^4}/(3\meanval{m^2}^2)$.
The latter are known to cross at a critical fixed point $U^{(4)}_*$ defining the true critical temperature,
yet, since the crossing point drifts smoothly over our range of lattice sizes, we assume
a power law of the form $L^w$ for $U^{(4)}(L)$, with an unknown exponent $w$ \cite{Binder1981}.
In addition, these two quantities are advantageously used to obtain critical temperatures
in the second-order regime as well (see Sec.~\ref{sec:second_order} for more details on this issue).

\begin{figure}[bt]
	\centering
	\InsertFig{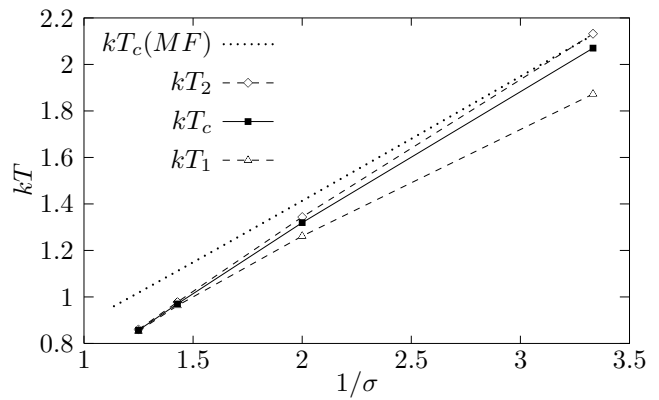} 
	\caption{Spinodal curve for $0.3 \leq \sigma \leq 0.8$ ($q=5$). The transition temperature $T_c$ 
	is indicated by filled squares, and the limits of metastability $T_1$ and $T_2$ by triangles and diamonds, respectively.
	Errors are smaller than the size of symbols, and lines are drawn to guide the eyes.
	The dotted line shows the transition temperature as predicted by MF theory. }
	\label{fig:spinodal_q_05}
\end{figure}

Results for all temperature estimates are summarized in Table~\ref{table:all_temperatures} for $q=3,5,7,9$, 
and sketched in Fig.~\ref{fig:spinodal_q_05} for $q=5$.
As expected according to FSS theory, both definitions of the transition
temperature, i.e., using equal peak weights vs equal peak heights, lead
within error bars to the same estimates  at infinite lattice size.
Other quantities $T_c(\chi)$ and $T_c(U^{(4)})$ yield very similar results, 
with a discrepancy never exceeding 1\%.

\begin{table*}
	\caption{Estimates of the critical temperature in the first- and second-order regimes (the latter is indicated by an asterisk):
		MF, mean-field;
		$\chi$, using location of peaks of the susceptibility;
		$U^{(4)}$ using crossing points of Binder cumulants of the magnetization;
		eqw,eqh, using the free energy, where $T_c$ corresponds to equal peak weights and heights, respectively;
		Ref. \cite{GlumacUzelac1997_1998}, MC study based on multihistogramming and the Luijten-Bl\"ote cluster algorithm ($q=3$) and a standard metropolis
			algorithm ($q=5$);
		Ref. \cite{Monroe1999}), cluster mean-field method combined with an extrapolation technique based on the VBS (Vanden Broeck and Schwartz) algorithm; 
		Ref. \cite{GlumacUzelac1993}), transfer matrix method combined with FRS.
	}
	\begin{ruledtabular}
\begin{tabular}{ll|l|llll|lll} 
$q$ & $\sigma$	& $T_c$ (MF) 	& $T_c(\chi)$ 	& $T_c(U^{(4)})$ & $T_c$(eqh)	& $T_c$(eqw)	& $T_c$ (Ref. \cite{GlumacUzelac1997_1998})	& $T_c$ (Ref. \cite{Monroe1999}) & $T_c$ (Ref. \cite{GlumacUzelac1993}) \\ \hline
%
3 & 0.2		& 4.034		& 3.97(1)	& 3.98(1)	& 3.94(1)	& 3.97(1)	& 3.70\footnotemark[1]	& 		& 3.7023 \\ 	 	
&   0.3		& 2.836		& 2.72(1)	& 2.72(1)	& 2.71(1)	& 2.71(1)	& 2.70\footnotemark[1]	& 2.71669	& 2.5893 \\ 
&   0.4		& 2.240		& 2.086(4)	& 2.089(6)	& 2.075(5)	& 2.074(4)	& 2.08\footnotemark[1]	&		& 2.0247 \\ 
&   0.5		& 1.884		& 1.691(3)	& 1.685(3)	& 1.686(4)	& 1.684(2)	& 1.70\footnotemark[1]	& 1.68542	& 1.6631 \\ 
&   0.6		& 1.649		& 1.44(1)	& 1.43(1)	& 1.43(1)	& 1.43(1)	& 1.41\footnotemark[1]	& 		& 1.4000 \\ 
&   0.7		& 1.482		& 1.196(3)	& 1.19(1)	& 1.18(1)	& 		& 1.19\footnotemark[2]	& 1.1968	& 1.1942 \\ 
&   0.8$^*$	& 1.358		& 1.019(4)	& 1.03(1)	&		&		& 1.01\footnotemark[2]	&		& 1.0231 \\ 
&   0.9$^*$	& 1.262		& 0.876		& 0.875		&		&		& 0.88\footnotemark[2]  & 0.8785	& 0.874  \\ 
\hline
%
5&  0.3		& 2.127		& 2.07(1)	& 2.07(1)	& 2.072(6)	& 2.070(4)	& 2.033\footnotemark[1]	& 2.06900	& 1.736 \\	 
&   0.5		& 1.413		& 1.321(3)	& 1.319(4)	& 1.319(3)	& 1.319(2)	& 1.297\footnotemark[1]	& 1.31638	& 1.245 \\
&   0.7		& 1.111		& 0.973(1)	& 0.973(2)	& 0.970(3)	& 0.970(2)	& 0.981\footnotemark[1]	& 0.96963	& 0.956 \\
&   0.8		& 1.018		& 0.854(1)	& 0.853(1) 	& 0.857(1)	& 0.857(1)	&			&		& 0.844	\\ 
&   0.9$^*$	& 0.947		& 0.743(2)	& 0.739(4)	& 		&		&			& 0.74673	& 0.745 \\ 
\hline
%
7&  0.2		& 2.600 	& 2.58(1) 	& 2.58(2)	& 2.578(2)	& 2.577(1)	&&& \\
&   0.4		& 1.444 	& 1.395(5) 	& 1.394(4)	& 1.394(1)	& 1.393(1)	&&& \\ 			 
&   0.6		& 1.063 	& 0.986(2) 	& 0.985(3)	& 0.984(1)	& 0.986(1)	&&& \\ 		 
&   0.8		& 0.875 	& 0.764(1) 	& 0.763(1) 	& 0.764(1)	& 0.764(1)	&&& \\ 
&   0.9		& 0.814		& 0.677(1)	& 0.676(1)	&		&		&&& \\ 
\hline 		
%
9&  0.2		& 2.353		& 2.33(1)	& 2.33(1)	& 2.33(1)	& 2.32(1)	&&&\\ 
&   0.3		& 1.655 	& 1.626(3)	& 1.625(4)	& 1.627(3)	& 1.626(1)	&&&\\ 
&   0.5		& 1.099 	& 1.052(2)	& 1.051(2)	& 1.050(3)	& 1.052(1)	&&&\\
&   0.7		& 0.864		& 0.793(2)	& 0.792(2)	& 0.794(2)	& 0.794(1)	&&&\\ 
&   0.8		& 0.792 	& 0.705(2) 	& 0.704(1)	& 0.704(1)	& 0.704(1)	&&&\\ 
\end{tabular}
	\end{ruledtabular}
	\footnotetext[1]{Refers to $1/K_e(\Delta F)$.}
	\footnotetext[2]{Refers to $1/K_e(U^{(4)})$.}
\label{table:all_temperatures}
\end{table*}

\begin{figure}[bt]
	\centering
	\InsertFig{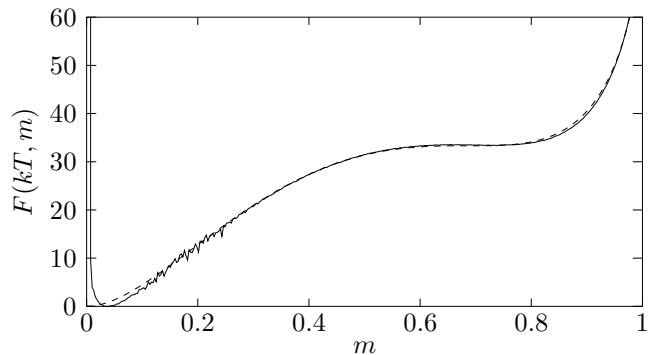}
	\caption{Partial free energy $F(kT,m)$ for $q=9,\sigma=0.3,L=400$ (solid line), together with the MF prediction (dashed line) as given
	by Eq.~(\ref{equ:free_energy_MF}).}\label{fig:free_energy_vs_MF}
\end{figure}

For all values of $q$, the transition temperatures progressively depart
from the MF line as $\sigma$ is increased. 
For $q=5$, for instance, the ratio between $T_c(\chi)$ and the MF value ranges from $97.3\%$ at $\sigma=0.3$ to $83.9\%$ at $\sigma=0.8$. 
We further notice that, for a given range of interaction, the adequacy of MF results is clearly improved at high $q$.
As illustrated in Fig.~\ref{fig:free_energy_vs_MF} for $q=9,\sigma=0.3$, and $L=400$, this agreement also holds, even
at finite lattice sizes, for the shape
of the partial free energy $F(kT,m)$ and the position of metastability plateaus.
For $q=3$ and $q=5$, we can readily compare the transition temperatures with earlier MC studies. 
Results obtained in \cite{GlumacUzelac1997_1998} using either the
Luijten-Bl\"ote cluster algorithm $(q=3)$ or
a standard metropolis algorithm ($q=5$) 
are in fairly good agreement with ours within an error bar that does not exceed 1\%, except in the
case $\sigma=0.2$, where our estimate lies much closer to the MF prediction.
We further compared our estimates with those obtained in \cite{Monroe1999} using a cluster mean-field method,
and in \cite{GlumacUzelac1993} using a transfer matrix approach.
As illustrated in Table~\ref{table:all_temperatures}, results
 obtained using the cluster mean-field approach combined with the VBS extrapolation algorithm yield a perfect match, 
 with a deviation as low as 0.1\% on average over the whole range of $\sigma$ values.
The discrepancy with estimates obtained using the transfer matrix method is slightly higher and amounts to 2\% on average,
except for low values of $\sigma$ where the agreement of our results with the MF prediction is, here again, far better. 

\begin{figure}[bt]
	\centering
	\InsertFig{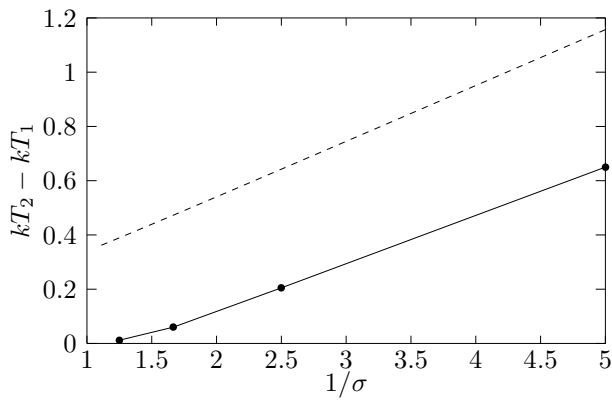}
	\caption{Difference between temperatures of metastability $dkT_m=kT_2-kT_1$ vs $1/\sigma$ for $q=7$ 
	(circles connected by solid lines).
	Errors are smaller than the size of symbols.
	MF prediction is shown for comparison (dashed line).}
	\label{fig:dTc_vs_inv_sigma}
\end{figure}

\begin{figure}[bt]
	\centering
	\InsertFig{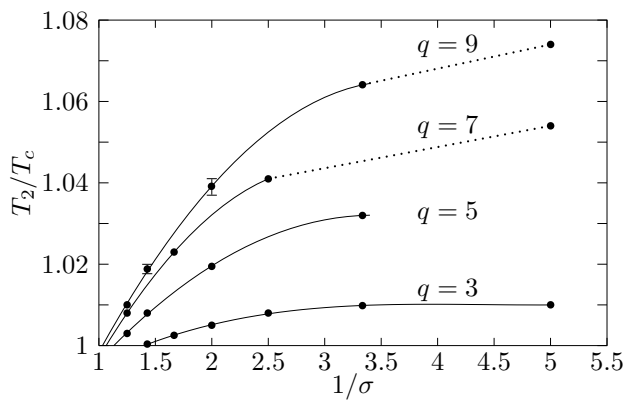}
	\caption{$T_2/T_c$ vs $1/\sigma$ for $q=3,5,7,9$. Solid lines indicate polynomial fits.
	Dotted lines are guides to the eyes. Error bars are smaller than the size of dots, except where explicitely indicated.}
	\label{fig:tmeta12_over_tc}
\end{figure}

\subsection{Change of regime} \label{sec:tricritical} 

\begin{figure}[bt]
	\centering
	\InsertFig{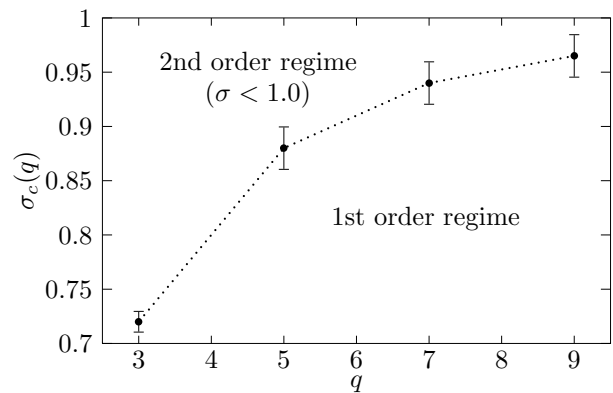}
	\caption{Phase diagram computed using FSS properties of spinodal points, for $\sigma<1.0$.
	Dotted lines are shown to guide the eyes.}
	\label{fig:phase_diagram}
\end{figure}

As can be viewed in Fig.~\ref{fig:spinodal_q_05}, spinodal points merge slightly above $\sigma\sim 0.8$ for $q=5$,
and this indeed signals a change of the nature of the transition. 
By plotting $dkT_m=kT_2-kT_1$ against $1/\sigma$, we observe that for all values of $q$ the
points fit quite well on a line for low enough $\sigma$, and the slope of this line tends toward that of 
the MF curve. The case $q=7$ is sketched in Fig.~\ref{fig:dTc_vs_inv_sigma}, where it is clear that the point at $\sigma=0.6$
marks the border between the linear and nonlinear behavior, illustrating the weakening of the first-order transition as $\sigma_c$ is approached.
Since temperatures appear to scale as $1/\sigma$ in the vicinity of the MF regime, 
it is thus more appropriate to work with $T_1/T_c$ and $T_2/T_c$,
for the scaling factors will then cancel out neatly except when approaching $\sigma_c(q)$. 
As mentioned above, the latter ratio, which is sketched in Fig.~\ref{fig:tmeta12_over_tc}, 
 offers a higher precision through a larger free energy plateau. 
 As $\sigma$ falls off to the MF regime, this ratio tends, within error bars, to the value 
predicted by the MF theory, i.e., $T_2/T_c=1.01, 1.037, 1.059, 1.077$ for
$q=3,5,7,9,$ respectively. On the leftmost side of the graph, we witness a sharp decrease of $T_2/T_c$ as $\sigma\rightarrow\sigma_c$.
This brings a quite reliable way of determining $\sigma_c(q)$ without much
ambiguity, as opposed to, e.g., methods using the interfacial free energy or Binder cumulants. 
By fitting data points to a polynomial of degree $2$ for $q=5,7,9$, and of degree $3$ for $q=3$, which turned out to
yield the lowest error, we obtained the following numerical estimates:
$$
\begin{array}[c]{lll}
q	&\ &	\sigma_c	\\ 
3	&\ &	0.72(1)	\\
5	&\ &	0.88(2)		\\
7	&\ &	0.94(2)		\\ 
9	&\ &	0.965(20)		\\ 
\end{array}
$$
The graph of $\sigma_c(q)$ is sketched in Fig.~\ref{fig:phase_diagram} for convenience.
Considering the global shape of this graph, it is reasonable to expect 
$\sigma_c(q)\rightarrow 1$ as $q\rightarrow\infty$. This would be clearly consistent with Cardy's scenario
(as mentioned in the Introduction), according to which the border case $\sigma=1.0$ corresponds to a KT-like transition governed by
topological defects. 

\subsection{Unexpected FSS behavior of correlation lengths and the dynamics of first-order transitions}

\begin{figure}[bt]
	\centering
	\InsertFig{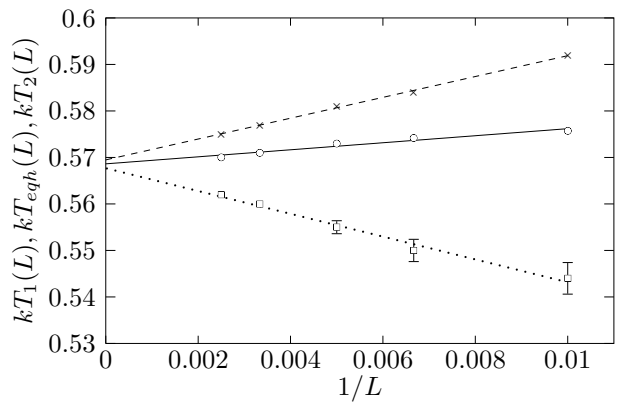}
	\caption{Linear fit of finite size temperatures vs $1/L$ for $q=9,\sigma=1.0$.
	Dotted, solid, and dashed lines correspond to $kT_1$, $kT_{eqh}$, and $kT_2$, respectively.
	Error bars are smaller than the size of symbols, except where explicitly indicated.
	In the limit $L\rightarrow\infty$,
	the difference between temperatures of metastability tends to $0.0012$. Within our error bars, the transition
	is thus clearly not of the first order.}
	\label{fig:critical_temperatures_pfe_q9s10}
\end{figure}

Let us now briefly inspect the case $q=9,\sigma=1.0$, where a simple analysis based on the shape of the free energy
 at a given lattice size might be markedly misleading. 
 In \cite{BayongDiepDotsenko1999}, a first-order transition for $q\geq 9$ was reported on the basis
of the observation of a double-peaked energy histogram. We have performed a series of simulations at $L=50,100,150,200,300,$ and $400$
for this set of parameters and computed corresponding (finite-size) spinodal temperatures $T_1(L)$ and $T_2(L)$ using the partial
free energy $F(kT,m)$. As may be noticed in Fig.~\ref{fig:critical_temperatures_pfe_q9s10},
a striking feature of this limiting case is the existence of metastable states at all finite lattice sizes, 
with a first-order character strongly enhanced at low sizes, despite the fact
that FSS theory yields $T_2-T_1=0$ in the thermodynamic limit.
It turns out that the transition is clearly not of the first order in
the thermodynamic limit, and this feature was also confirmed for $q=6,7,$ and $8$; for $q<6$, a precise location of
metastable states became impracticable.

At first blush this behavior significantly contradicts the expected picture,
whereby for first-order transitions, the correlation length is finite and roughly independent of
the lattice size, and is merely connected to the size of clusters. As a result,
such transitions appear as if they were continuous until the lattice size 
overtakes the correlation length. With regard to SR models, this has been the standard scenario thus far, 
yet we feel strongly that this scenario may be somewhat
challenged, at least qualitatively to begin with,
 where models incorporating LR interactions are concerned.
 
To set the stage for an attempt to interpret this behavior, we first turn to the consequences of finite
lattice size on long-wavelength fluctuations when simulating LR models with
algebraically decaying interactions. The key point in the following discussion
is the nature of the phase transition as \textit{observed} from numerical data obtained 
at finite lattice size.
On a lattice of size $L$ with periodic boundary conditions, the largest allowed distance
between any two spins is $L/2$, and this also corresponds to the smallest interacting potential
affordable on a given lattice.  It is obvious that these spins experience a
stronger interacting potential whenever $L$ is small, and hence the whole array of spins
may be rigidly tied to an adequate extent for an order-disorder transition to
occur through metastability. When increasing the lattice size, on the contrary, spins being a distance
$L/2$ apart now experience weaker interaction, and this results in a softening of the
transition. Whether this softening might be sufficient to yield a change of
nature of the transition at some (either finite or infinite) lattice size, so that the transition may be 
 continuous in the thermodynamic limit, is however an unsettled question;
this assumption is borne out at least for $q=9$ and $\sigma=1.0$, as witnessed by our results.
Alternatively, we may say that the truncation of LR interactions at small
lattice size artificially shifts the model toward the MF regime, 
since the interacting potential now varies smoothly over the available distance of interaction.

As seems obvious to us, the usual physical meaning attributed to the correlation length
in the case of SR models, i.e., roughly speaking the average 
size of a cluster of contiguous spins having the same value, may no longer hold 
in the case of LR models: since all the spins of the lattice,
however distant they may be, are tied together through an interacting potential,
there is basically no need of a long-range order for two distant spins to  
already have slightly correlated fluctuations. In the context of first-order transitions,
this means that 
either clusters may extend well beyond the size permitted by the
value of the correlation length, 
or the correlation length itself may become infinite in the thermodynamic limit.
This behavior has indeed already been reported
 in models of DNA thermal denaturation \cite{Theodorakopoulos2000} as well as 
in the context of wetting \cite{PrivmanSvrakic1988}. 
 
In addition, we have performed simulations in the first-order regime at the finite-size transition
temperature. We used, however, a Metropolis
algorithm, since the associated dynamics is closer to the real nucleation or 
spinodal decomposition picture than with a multicanonical algorithm.
We observed indeed that clusters in the ordered phase always spanned the entire lattice, 
whatever the lattice size. As soon as the dynamics jumps from the disordered to
the ordered phase, which we monitored by comparing the energy with
the location of reweighted energy histogram peaks, a single cluster forms very rapidly and
nudges its way through the crowd of disordered spins so that it swiftly occupies the
whole lattice. Thus, if both phases coexist insofar as, e.g., the energy histogram has
 a double-peaked structure, they actually do not coexist at the same time and
 merely alternate in time, as opposed to what is considered the usual SR picture.
 In this respect, we would like to raise some challenging question
 regarding the dynamics of first-order transitions in the LR case: 
 (i) Since both phases do not coexist at the same time, what physical meaning
 should be given to the interfacial free energy?
 (ii) Does a mechanism similar to nucleation
 take place in a LR system, and if in the affirmative, how can it be reconciled  
 with the mechanism of cluster growth involved in SR models?

\subsection{Beyond the tricritical line: From LR to SR behavior} \label{sec:second_order}

\begin{figure}[bt]
	\centering
	\InsertFig{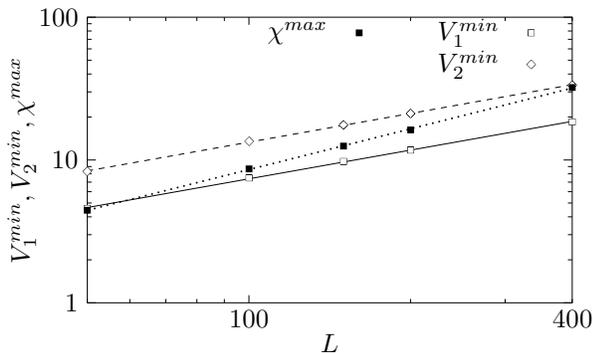}
	\caption{Fit of $V_1^{min}$, $V_2^{min}$, and $\chi^{max}$   
	vs $L$ on a log-log scale, for $L=50,100,150,200,400$ ($q=5, \sigma=0.9$).
	Errors are smaller than the size of symbols.}
	\label{fig:critical_exponents_q5s09} 
\end{figure}

We now focus on some critical properties in the second-order regime $\sigma_c(q) < \sigma < 1.0$,
then we investigate the crossover from LR to SR behavior.  
As mentioned in \cite{LuijtenBlote1997}, "standard" FSS theory is valid for LR systems provided the  
effective upper critical dimension $d^*=2\sigma$ is greater than the geometrical dimension $d=1$, i.e.,
$\sigma>0.5$. Thus for $q\geq 3$ we assume "standard" finite-size scaling
equations to be valid. We first determine the critical exponent $\nu$ using
 $n$th-order cumulants of the magnetization, i.e.,
$V_n=d\ln\meanval{m^n}/d\beta$, which
have minima obeying the scaling law $V_n^{min} \propto L^{1/\nu}$ \cite{FerrenbergLandau1991}.
Our approach is to compute two numerical estimates of $\nu$ by 
fitting reweighted averages of $V_1^{min}$ and $V_2^{min}$ to a power law of the lattice size,
and then to average over both values.
Other critical exponents, i.e., $\beta$ and $\gamma$, are computed using similar scaling 
laws, i.e., $M(T_c(\infty)) \propto L^{-\beta/\nu}$, 
and $\chi^{max} \propto L^{\gamma/\nu}$.
Figure~\ref{fig:critical_exponents_q5s09} shows a power-law fit of peaks of $V_1$, $V_2$ and $\chi$ against
the lattice size obtained for $q=5,\sigma=0.9$. Points lie neatly on a straight line when using a log-log
scale, and give the following estimates: $1/\nu_1=0.668(2)$, $1/\nu_2=0.669(2), \gamma/\nu=0.940(4)$. 
Error bars were computed using a bootstrap procedure.
Once $\nu$ is known, we fit  finite-size temperatures $T_{c}(L)$  defined from peaks of the 
magnetic susceptibility to a power law of the form 
$T_c(L) = T_c(\infty) + \lambda L^{-1/\nu}$ and obtain an estimate of the critical temperature.
With regard to critical couplings obtained from Binder cumulants of the magnetization, we follow the same procedure
as in the first-order regime.
Finally, the critical exponent $\beta$ is determined by fitting $M(T_c(\infty))$ to
a power law of the lattice size,
and slowly varying the temperature at which $M$ is to be sampled
until the best fit is obtained. In the example considered above,
 this leads to $\beta/\nu=0.103(2)$.
 Results for other pairs of $(q,\sigma)$ values are summarized in 
Table~\ref{table:all_temperatures} and Table~\ref{table:exponents_allq}.
For the borderline case $\sigma=1.0$, only exponent ratios are shown.
It can be seen that our estimates match fairly well those obtained from a previous MC study \cite{BayongDiepDotsenko1999}, 
and that the discrepancy with results obtained from a transfer matrix approach in \cite{GlumacUzelac1993} never exceeds $8\%$.
As opposed to the conjecture made in \cite{CannasMagalhaes1997}, the exponent $\nu$ does clearly depend on $q$.

\begin{table}[bt]
	\caption{Critical exponents in the second-order regime $\sigma>\sigma_c(q)$, and $q=3,4,5$. 
	Shown for comparison are results from Ref.~\cite{GlumacUzelac1993} 
	obtained using a transfer matrix method, and from Ref.~\cite{BayongDiepDotsenko1999} using a MC histogram approach.}
	\begin{ruledtabular}
	  \begin{tabular}{llllll} 
	    $q$ & $\sigma$	& $\nu^{-1}$	& $\nu^{-1}$ \cite{GlumacUzelac1993} & $\gamma/\nu$	& $\beta/\nu$	 \\ \hline 	
	    3  & 0.8	& 0.624(6)	& 0.574		& 0.842(5)	& 0.101(5)	\\ 
	       & 0.9	& 0.54(1)	& 0.491		& 0.908(5)	& 0.053(5)	\\ 
	       & 1.0	&		&		& 0.96(1)	& 0.025(8)	\\ 	   
	    4   & 0.8	& 0.71(1)	& 0.67		& 0.882(3)	& 0.122(4)	\\ 
	        & 0.9	& 0.610(5)	& 0.56		& 0.920(4)	& 0.050(3)	\\ 
	        & 1.0	&		&		& 0.96(1)	& 0.022(9)	\\ 	   	 
	    5  & 0.9	& 0.668(2)	& 0.62		& 0.940(4)	& 0.103(2)	\\ 
	       & 1.0 	&		&		& 0.97(1)	& 0.04(1)	\\ 
	       & 1.0 \cite{BayongDiepDotsenko1999} & & & 0.966 	& 0.017 	\\ 
	  \end{tabular}
	\end{ruledtabular}
	\label{table:exponents_allq}
\end{table}

If the relation $\sigma = 2 - \eta$ derived in \cite{FisherMaNickel1972} is indeed exact for $q \geq 3$, we should
thus observe the simple behavior $\gamma/\nu = 2-\eta=\sigma$ in the second-order regime. 
As illustrated in the fifth column of Table~\ref{table:exponents_allq},
the qualitative behavior follows the conjecture, yet clearly $\sigma<2-\eta$, and
the discrepancy is remarkably higher for $q=5$ than for $q=3$. Moreover, while it appears to shrink to $0$ as $\sigma\rightarrow 1$,
 it is unclear whether $\gamma/\nu$ varies linearly with $\sigma$, considering the small number of points available.

In order to get a deeper insight into the crossover to the SR regime,
we then conducted several simulations at $q=3$ for $\sigma$ above the borderline value $\sigma_{co}=1$.
This value has been reported to play the role of a critical range of interaction beyond which a crossover from LR
to SR behavior sets in. According to \cite{Sak1973, TheumannGusmao1985}, $\sigma_{co}=2-\eta_{SR}$, where
$\eta_{SR}$ denotes the value of the $\eta$ exponent in the SR case.
Since $\gamma/\nu=1$ for all values of $q$ in the SR case, $\eta_{SR}=1$, and this indeed leads to $\sigma_{co}=1$.
It should be noted, however, that this definition,  
as initially proposed by Sak in \cite{Sak1973} on theoretical grounds,
as well as the exact location of $\sigma_{co}$ within the interval $[1.0,2.0]$, is still controversial.
As shown in Table~\ref{table:exponents_allq}, $\gamma/\nu$ indeed appears to reach its SR value as $\sigma \rightarrow 1^-$,
yet this ratio proves no longer reliable above the borderline value, as we will witness in a moment, and
reliance on other quantities becomes necessary.

We first review some exact results concerning the SR regime, which we obtained using an exact transfer matrix
method. For $q=3$, the transfer matrix is a $3 \times 3$ matrix having three eigenvalues, which in zero external field
read $\lambda_1=3\cosh (\beta/2) - \sinh (\beta/2)$, $\lambda_2=\lambda_3=2 \sinh (\beta/2)$, where $\beta=1/kT$. 
By retaining the largest eigenvalue $\lambda_1$ only, and taking the limit $L\rightarrow\infty$, 
we successively obtain the free energy per spin
$$
F(\beta)= - \frac{\ln(2+e^{\beta})}{\beta}
$$
and the specific heat
$$
C_v(\beta) = \frac{2 \beta^2}{\left(\sinh \beta/2 - 3 \cosh \beta/2\right)^2}
$$
From there on, the correlation length is then computed using the standard formula 
$\xi = 1/\ln(\lambda_1/\lambda_2)$, which then yields
$$
	\xi(\beta) = \left[\ln\frac{3\coth \beta/2 -1}{2}\right]^{-1}
$$
Finally, the magnetic susceptibility is obtained using the fluctuation-dissipation relation, which gives
$$
	\chi(\beta) = \frac{8}{27} \beta (1 + 2 e^{\beta})
$$
It is then straightforward to show that $\lim_{\beta\rightarrow \infty} \ln\chi(\beta) / \ln\xi(\beta) = \gamma/\nu =1$.
However, evaluating this ratio at finite inverse temperature, i.e., for a finite correlation length as imposed by
a finite lattice size, yields a greatly overestimated result. For instance,  we obtain
$\gamma/\nu \sim 1.3$ for $L=400$, a feature which is supported by our simulation results, 
e.g., $\gamma/\nu=1.02(1)$, $1.14(1)$, and $1.23(1)$ for $\sigma=1.1$, $1.5$, and $4.0$, respectively. 
Since the last two values are clearly overestimated, this in effect indicates the presence of exponential divergences and 
as a by-product drastically slow convergence of the correction to scaling.   

\begin{figure}
	\centering
	\InsertFig{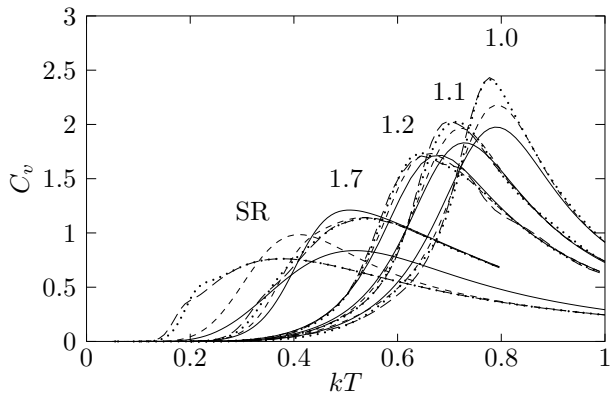}
	\caption{Specific heat for various lattice sizes and 
	$\sigma=1.0,1.1,1.2,1.7$ and the pure SR case (obtained using a transfer matrix method), from right to left.
	Data for other values of $\sigma$ have been omitted in order to preserve the clarity of the figure.
	$C_v$ was computed using the fluctuation-dissipation relation 
	$C_v=(\langle E^2 \rangle - \langle E \rangle^2)/ (kT^2 L)$.
	Solid, dashed, dotted and long-dashed styles refer to $L=50,100,200$, and $400$ respectively,
	except for the SR case where they refer to $L=5,10,100,200$.
	}
	\label{fig:cv_allsigma_vs_SR}
\end{figure}

This analysis was corroborated by a study of the shape of the specific heat, 
which turns out to provide the most tractable approach at 
medium lattice sizes where distinguishing between the SR and the LR regime is concerned. 
In the thermodynamic limit, $C_v(\beta)$ admits a maximum $C_v^{max}=0.7618$ at $kT_m=0.3767$.
It is enlightening to investigate the nonmonotonic behavior of this maximum at finite $L$, and
this may be carried out by computing $F(\beta,L)$ and then $C_v(\beta,L)$ while
retaining all three eigenvalues. Since the calculation is fairly involved, and the final result admits no simple
expression, we shall hereafter simply refer to the corresponding curve sketched in Fig.~\ref{fig:cv_allsigma_vs_SR}. 
When $L$ is increased, the peak of the specific heat first increases to a maximum,
and then graphs of $C_v$  collapse and merge gently as 
the thermodynamic limit is approached. Whenever it is
witnessed in graphs obtained from simulation data, this feature thus signals a SR-like behavior.

\begin{figure}
	\centering
	\InsertFig{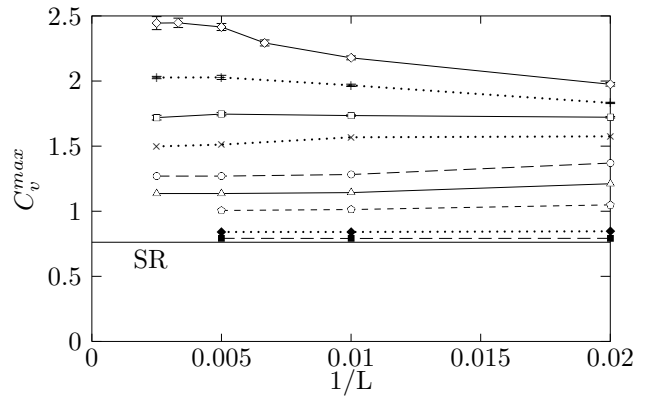}
	\caption{Maximum of the specific heat vs inverse lattice size for $\sigma=1.0,1.1,1.2,1.3,1.5,1.7,2.0,3.0,4.0$
	from top to bottom. The solid line is a reminder for the
	(exact) SR case in the thermodynamic limit. Other lines are guides to the eyes.}
	\label{fig:cvmax_allsigma_vs_SR}
\end{figure}

\begin{figure}
	\centering
	\InsertFig{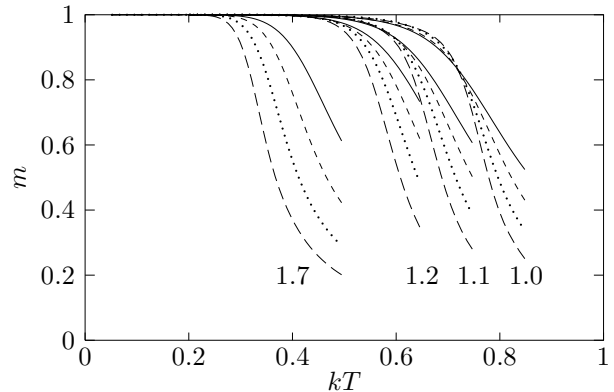}
	\caption{Magnetization vs $kT$ for $\sigma=1.0, 1.1, 1.2$, and $1.7$ from right to left.
	Solid, dashed, dotted and long-dashed styles refer to $L=50,100,200$, and $400$ respectively.
	}
	\label{fig:mag_all_sigma}
\end{figure}

Simulations were performed for
$1.0 \leq \sigma \leq 4.0$ for various lattice sizes between
$L=50$ and $L=400$, and we set the initial canonical temperature to $kT_0=1.0$ so that
the maximum of $C_v$ would be clearly visible within the whole range $\sigma \geq 1.0$.
As appears obvious from a glance at Fig.~\ref{fig:cv_allsigma_vs_SR}, the cases $\sigma=1.0$ and
$\sigma=1.1$, on the one hand, and $\sigma \geq 1.2$, on the other hand, display fairly distinct
qualitative behaviors. For $\sigma=1.0$, the specific heat reaches its maximum monotically, at least
for the lattice sizes that were investigated. The slowing down in the increase rate as $1/L\rightarrow 0$ 
allows one to assess a finite maximum in the thermodynamic limit, and this clearly shows that $C_v$ is a nondivergent
quantity, thus bringing support to Cardy's scenario whereby the transition has a KT-like nature on the borderline $\sigma=1.0$.
The same behavior is observed for $\sigma=1.1$.
On the contrary, the qualitative behavior is clearly different for $\sigma \geq 1.2$, where 
the maximum of $C_v$ first decreases with increasing lattice size, and then quickly reaches a plateau reminiscent
of the exact SR behavior investigated above. While this plateau only slowly reaches the exact SR value as 
$\sigma \rightarrow 4.0$ (see Fig.~\ref{fig:cvmax_allsigma_vs_SR}),
 we can however conclude that the behavior is already SR-like.   
This assertion can be further confirmed by considering the magnetization, as 
sketched in Fig.~\ref{fig:mag_all_sigma}. Graphs of this quantity clearly merge slightly above $m=0$,
whenever $\sigma \geq 1.2$; hence there is no transition at finite temperature.
 While for $\sigma=1.1$ there remains some ambiguity due to statistical errors, for $\sigma=1.0$ the curves now clearly intersect  
 around $kT \sim 0.7$, which at least shows that the behavior is no longer SR-like. 
We finally compute critical temperatures  from the crossing points of Binder cumulants of the magnetization. We obtain
$\beta_c=3.3$, $6.5$, and $19$ for $\sigma=1.1$, $1.3$, and $1.5$. As for $\sigma=1.7$ and $\sigma=2.0$, cumulants
 no longer cross except at $kT=0$ within
statistical error (the latter case yielding $\beta_c$ between $150$ and $200$, yet with excessive dispersion). 
While the crossover appears to take place in the very vicinity of the borderline
$\sigma=1.0$, the critical temperature actually dies off quite slowly to $0$ as $\sigma$ increases. 

All these numerical results lend support to Sak's scenario for $\sigma > 1.0$, namely, that
 a crossover from LR to SR behavior occurs whenever  $\sigma_{co}=2-\eta_{SR}$. Nonetheless, it is worth
 mentioning that we found this crossover to occur within the finite, yet narrow range 
$1.0 < \sigma < 1.2$, and the pure SR case to be reached in the limit $\sigma\rightarrow\infty$ only.
We feel strongly that this is consistent with the RG scenario of Theumann and Gusmao
\cite{TheumannGusmao1985}, whereby the crossover actually results from 
a competition between SR and LR fixed points. This competition, as seems obvious to us, may not resolve instantly 
whenever $\sigma$ crosses the borderline, and may thus blur this borderline over some finite region.     
 
\section{Conclusion} 

We have studied some critical properties of the long-ranged Potts model using a multicanonical implementation
of generalized ensemble algorithms. 
Our implementation of the iteration procedure needed to obtain the density of states
was shown to yield satisfying estimates of this quantity over a large range of energy and with much quicker 
and more stable convergence than with the initial historical algorithm.
The multicanonical algorithm allows one to efficiently circumvent the slowing down traditionally experienced
at first-order transitions, and at the same time makes the reweighting approach a fairly straightforward way
of examining thermodynamic quantities over a large range of temperature with strikingly modest numerical effort, i.e.,
by simulating over medium lattice sizes and performing a single long simulation run.
We have used this multicanonical approach to locate spinodal points in the first-order regime over a large range
of $q$ and $\sigma$ parameters. The shape of the spinodal curve in the vicinity of the change of regime
then yielded precise estimates of the tricritical value $\sigma_c(q)$ up to two digits.
In particular, the value $\sigma_c(3)=0.72(1)$ is perfectly consistent with the lower bound of $0.7$ proposed 
by Krech and Luijten \cite{KrechLuijten2000},
yet in terms of precision this is markedly better by an order of magnitude.
In this respect, our multicanonical implementation allows us to obtain numerical results whose accuracy is
at least comparable to that of previous numerical studies based on multihistogramming and
the LR cluster algorithm, although our simulations were performed on lattices having fewer than $400$ spins.
We feel strongly that this approach might be successfully applied to other spin models incorporating LR
interactions, e.g., continuous spin models or frustrated systems. 

In addition, our study significantly extends the range of available estimates of critical couplings and exponents.
In the first-order regime, the agreement with MF predictions, and in particular with
Tsallis's conjecture $T_c \sim 1/\sigma$ in the limit $\sigma\rightarrow 0$ \cite{Tsallis1995}, is exceptionally good. 
In the second-order regime, the relation $\eta=2-\sigma$, conjectured to be exact for $q=2$, is shown to yield an increasingly
high discrepancy when $q$ is increased, and its validity may just be reinforced in the vicinity of $\sigma=1.0$.
We found however that the crossover from the LR to the SR regime occurs between $\sigma=1.0$ and $\sigma=1.2$, thus
lending strong support to Sak's conjecture.
Our detailed FSS analysis of the case $q=9,\sigma=1.0$ yielded one of the most surprising results of this study,
namely, the unexpected behavior of correlation lengths whereby the transition appears to be of the first order at finite lattice
size, despite the fact that FSS theory predicts a continuous transition in the thermodynamic limit.
We feel strongly that this may be accounted for by the truncation of the LR potential, which artificially brings
the model closer to the MF regime, yet we also pointed out that the physical meaning of the correlation length should
be somewhat challenged in the case of LR models.
The exact nature of the transition in the borderline case $\sigma=1.0$, however,
needs further investigation, especially at large $q$ where no results have been made available thus far. 
In this view, an efficient combination of a global update scheme with a multicanonical
approach would be of prior importance to reach far higher lattice sizes.

\begin{acknowledgments}
S.R. is indebted to Dr. A. Revel for providing access to computing facilities of the ETIS (ENSEA/UCP/CNRS, UMR 8051) research team.
"Laboratoire de Physique Th\'eorique et Mod\'elisation" is associated with CNRS (UMR 8089).
\end{acknowledgments}

\end{document}